\documentclass[prb,aps,
twocolumn,
groupedaddress,floats,showpacs,final,superscriptaddress,10pt]{revtex4-2}
\usepackage[latin3]{inputenc}
\usepackage[makeroom]{cancel}
\usepackage{graphicx}
\usepackage{amsmath}
\usepackage{amsfonts}
\usepackage{amssymb}
\usepackage{chngcntr}
\usepackage{color}
\usepackage{xcolor}
\usepackage{cancel}
\usepackage[left=2cm,right=2cm,top=2cm,bottom=2cm]{geometry}
\usepackage{caption}
\usepackage{subcaption}
\usepackage{microtype}
\usepackage{multirow}
\usepackage{booktabs}

\usepackage{graphicx}
\usepackage{dcolumn}
\usepackage{bm}
\usepackage{simplewick}
\usepackage{array}
\usepackage{appendix}
\usepackage{placeins}
\usepackage{subcaption}

\usepackage[normalem]{ulem}

\definecolor{purple}{rgb}{0.5,0,0.6}

\RequirePackage[
   hyperindex,colorlinks,bookmarksnumbered,
   plainpages=true,pdfstartview=FitH]{hyperref}
\hypersetup{linkcolor=blue,urlcolor=blue,citecolor=blue}

\begin{document}

\title{Resonant nonlinear THz valley Hall effect in two-dimensional Dirac semiconductors}




\author{V. N.~Ivanova}
\affiliation{Technion -- Israel Institute of Technology, Haifa, 3200003, Israel}
\affiliation{Guangdong Technion -- Israel Institute of Technology, 241 Daxue Road, Shantou, Guangdong, China, 515063}


\author{V.~M.~Kovalev}
\affiliation{Novosibirsk State Technical University, Novosibirsk 630073, Russia}

\author{O.~V.~Kibis}
\affiliation{Novosibirsk State Technical University, Novosibirsk 630073, Russia}

\author{I.~G.~Savenko}
\email[ivan.g.savenko@gmail.com]{}
\affiliation{Guangdong Technion -- Israel Institute of Technology, 241 Daxue Road, Shantou, Guangdong, China, 515063}
\affiliation{Technion -- Israel Institute of Technology, Haifa, 3200003, Israel}

\date{\today}

\begin{abstract}
We predict a cyclotron resonance enhancement in the nonlinear valley Hall response of inversion-asymmetric two-dimensional semiconductors subjected to crossed terahertz electric and static magnetic fields. 
By employing a two-band Hamiltonian that captures the essential orbital texture of the electronic states and broken inversion symmetry, we develop a kinetic theory that accounts for antisymmetric skew scattering from impurities and the trigonal warping of the electron dispersion valleys. 
Solving the Boltzmann transport equation, we uncover resonant photocurrents that exhibit a sharp, polarity-switching cyclotron peak and a nontrivial polarization response dictated by the underlying D$_{3h}$ crystal symmetry. 
Our results establish a universal mechanism for frequency-selective, phase-sensitive control of valley current that is directly accessible in monolayer transition-metal dichalcogenides. 
This work provides a pathway for harnessing resonant nonlinear transport in valleytronic and terahertz optoelectronic devices.
\end{abstract}

\maketitle

\section{Introduction}
Nonlinear Hall and photogalvanic effects~\cite{Sturman1992, Ivchenko2005, Ganichev2006} in two-dimensional (2D) materials have recently emerged as powerful spectroscopic probes of Berry curvature, orbital magnetism, and inversion-symmetry breaking~\cite{das2024nonlinear, wang2024nonlinear, bandyopadhyay2024non, Sodemann2015}. 
In electronic systems lacking inversion symmetry, various features of transport phenomena can exist~\cite{LAWTON2002699, Kibis_1999, Kibis_2001, Kibis_2002}. 
In particular, skew scattering on impurities, an extrinsic mechanism arising from asymmetric scattering amplitudes, gives rise to transverse currents quadratic in the driving electric field~\cite{glazov2020skew}. Moreover, when a perpendicular static magnetic field is applied, the cyclotron motion of carriers can produce a resonant enhancement of the nonlinear response~\cite{huang2021resonant}
as demonstrated by the observation of the cyclotron-resonance-assisted photon drag effect in InSb-based quantum wells~\cite{PhysRevB.89.115435}.

The cyclotron resonance and the nonlinear valley Hall effect in monolayer transition metal dichalcogenides (TMDs) such as $\mathrm{MoS}_2$ and $\mathrm{WSe}_2$ are of central importance for the development of valleytronics, spintronics, and THz optoelectronic devices~\cite{min2023strong, huang2023giant}. 
These phenomena exploit the valley degree of freedom, in which carriers in the $K$ and $K'$ valleys possess opposite Berry curvatures, enabling highly selective valley transport. 

The linear valley Hall effect was first experimentally demonstrated in $\mathrm{MoS}_2$ transistors by Mak et al.~\cite{mak2014valley}, who achieved valley-selective carrier population via circularly polarized light and observed a light-helicity-controlled anomalous Hall voltage. Building upon this foundation, Wu et al.~\cite{wojciechowska2024intrinsic} studied the intrinsic character of the valley Hall transport in atomically thin $\mathrm{MoS}_2$ without external symmetry breaking, reporting a characteristic cubic scaling of nonlocal resistance with local resistance that remains observable even at room temperature.

However, in realistic (disordered) 2D semiconductors, extrinsic scattering channels usually play the dominant role~\cite{glazov2020skew}. 
The first-principles calculations reveal that the skew scattering on vacancies dominates the valley Hall conductivity in monolayer $\mathrm{MoS}_2$ at low impurity concentrations, leading to a divergent conductivity in the clean limit~\cite{liu2024dominance}. Analogous conclusions can be drawn not only for electrons, but also for excitons in semiconductors ~\cite{PhysRevLett.125.157403}. 
Furthermore, disorder-induced nonlinear Hall effects that preserve time-reversal symmetry have been systematically analysed~\cite{du2019disorder, du2021quantum}.

The field has rapidly progressed in the nonlinear regime, with the introduction of the nonlinear valley Hall effect as a second-order response capable of generating valley currents even in systems that simultaneously preserve inversion and time-reversal symmetries~\cite{das2024nonlinear}.
This concept has since been extended to bilayer TMDs, twisted bilayer $\mathrm{WSe}_2$ exhibiting giant nonlinear Hall signals, and strain-engineered monolayers~\cite{okyay2025unconventional, xiong2025strain}. 
Furthermore, resonant nonlinear Hall phenomena under perpendicular magnetic fields were predicted, and the cyclotron resonance itself was experimentally detected in $\mathrm{graphene/MoS}_2$ van der Waals interfaces via photo-induced thermionic emission~\cite{lee2021ultrafast, PhysRevB.87.235439} and bilayer graphene where resonantly excited edge photocurrents were observed~\cite{PhysRevB.103.125408, PhysRevB.105.155416}. 
Very recent experiments have further reported giant gate-tunable nonlinear valley Hall effects~\cite{yin2022tunable} and strong room-temperature bulk nonlinear Hall responses in spin-valley locked Dirac materials, underscoring the technological potential of frequency-selective valley transport~\cite{PhysRevB.92.165314, PhysRevMaterials.3.054205}.

The influence of a static magnetic field on the second-order response in 3D cubic materials has been extensively studied~\cite{Blokh1980, Ivchenko1984}.
Specifically, the photogalvanic effect (PGE) near the cyclotron resonance was investigated for the intraband regime, with the carrier scattering determined by dipolar impurities~\cite{Blokh1980}. 
This analysis was subsequently extended to include magnetic-field-induced PGE transport arising from interband transitions~\cite{Ivchenko1984, PhysRevB.90.205415}.
While the impact of magnetic fields on low-dimensional structures has been addressed~\cite{deyo2009semiclassicaltheoryphotogalvaniceffect}, prior studies remain confined to the classically low magnetic fields~\cite{Olbrich_2009, Diehl_2007, Bel_kov_2005, PhysRevB.90.205415, PhysRevB.93.075306}.
Other findings include THz photocurrent magneto-oscillations in HgTe \cite{PhysRevB.90.205415} and the theoretical prediction of a giant quantum ratchet effect under cyclotron resonance \cite{PhysRevB.93.075306}.
Moreover, a unified theory that would capture the resonant interplay between cyclotron motion, skew scattering, electron dispersion trigonal warping, and the nonlinear valley Hall response in inversion-asymmetric 2D semiconductors, particularly one that goes beyond the low-field limit and accounts for the full frequency-dependent structure of the photocurrent, has remained elusive.

In this Letter, we theoretically study the nonlinear valley Hall effect in 2D semiconductors exposed to an AC electric field and a static magnetic field. 
\begin{figure}[t!]
\includegraphics[width=0.99\columnwidth]{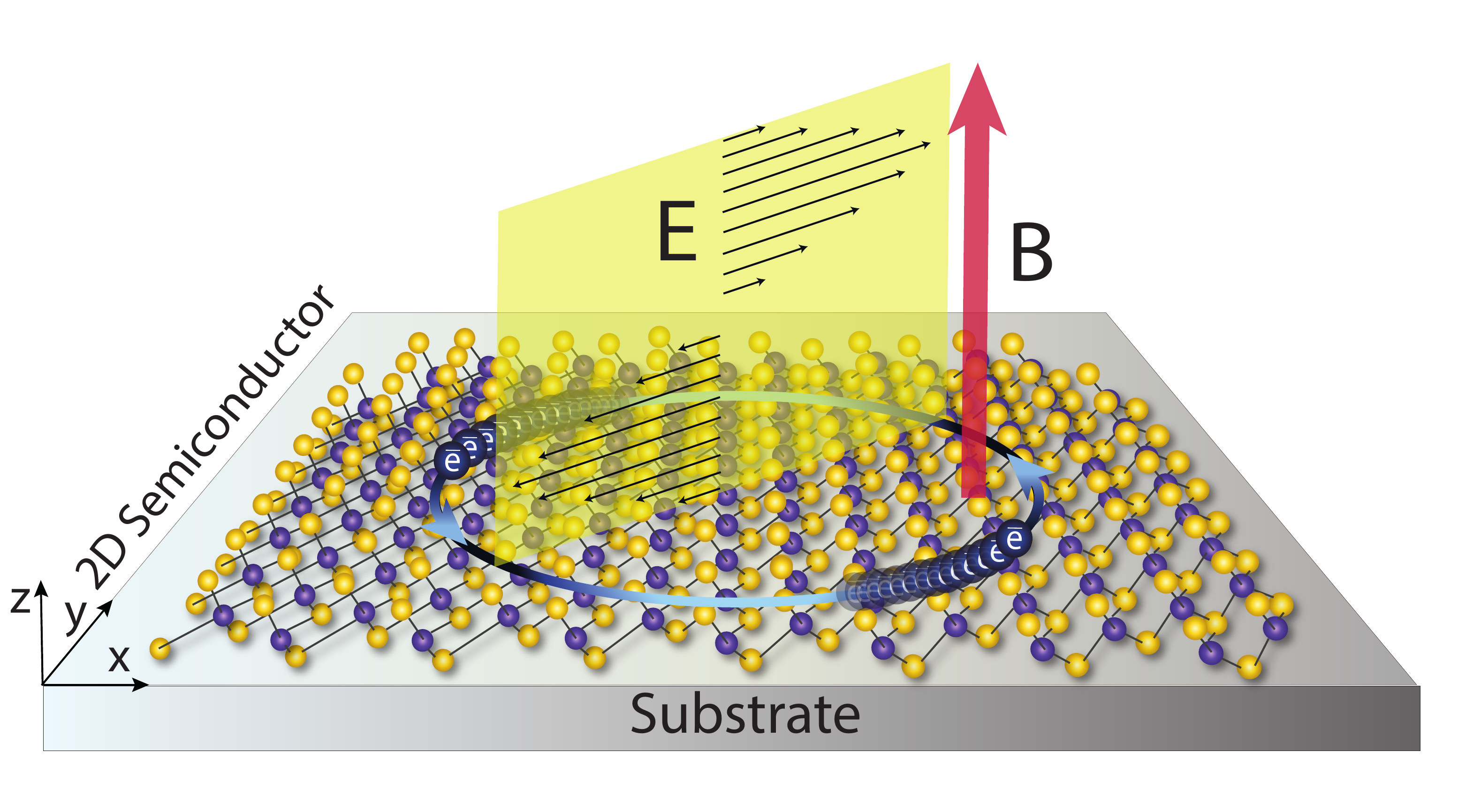} 
\captionsetup{justification=raggedright,singlelinecheck=false}
\caption{System schematic: a 2D semiconductor on a substrate subjected to a linearly polarized EM field ${\bf E}=(E_x, E_y, 0)=E_0(\cos\theta,\sin\theta,0)$, where $\theta$ is a polarization angle, and a normal permanent magnetic field $\mathbf{B}=(0,0,B)$.}
\label{Fig_currents}
\end{figure}
A phenomenological expression describing DC individual-valley electric current density in a 2D semiconductor possessing the $C_{3h}$ point group reads as
\begin{gather}\label{maineq}
j^{\eta}_x=\chi_{\eta}(|E_x|^2-|E_y|^2)+\lambda_{\eta}(E_x^*E_y+E_xE_y^*),\\\nonumber
j^{\eta}_y=-\chi_{\eta}(E_x^*E_y+E_xE_y^*)+\lambda_{\eta} (|E_x|^2-|E_y|^2).
\end{gather}
Thus, the valley ($\eta=\pm1$) response is characterized by two parameters, $\chi_{\eta}$ and $\lambda_{\eta}$, which represent functions of magnetic field. 
The influence of the magnetic field is twofold. 
First, it modifies the 
$\chi_{\eta} = \chi_{\eta}(\omega_c)$ coefficient, which describes the nonlinear response 
in the absence of a magnetic field, thereby introducing a dependence on 
the magnetic field strength. 
Second, it produces an additional contribution 
determined by the factor $\lambda_{\eta} = \lambda_{\eta}(\omega_c)$. 
Since both the coefficients 
depend on the cyclotron frequency $\omega_c$, they may exhibit cyclotron 
resonance behavior as the external electromagnetic field frequency $\omega$ approaches $\omega_c$. 
Below, we will focus on calculations of electric current in a single valley. 
The valley current, by definition, is the difference between currents in different valleys.

The generation of valley Hall electric current can be due to several microscopic mechanisms~\cite{glazov2020skew}. 
However, asymmetric skew scattering and trigonal warping of the electron dispersion in the valleys are usually the dominant processes.
Thus, we treat them here as such. 
To satisfy the cyclotron resonance conditions, these processes must be accompanied by a rigorous analysis of carrier motion in classical strong magnetic fields. 
Below, using an effective two-band Hamiltonian with linear and quadratic momentum terms that reflect the specific trigonal symmetry of the TMD materials, we solve the semiclassical Boltzmann kinetic equation to second order in the electric field, accounting for the strong (classical) magnetic field, and study the electric current density.


\section{Model and eigenstates}
An effective low-energy two-band Hamiltonian describing the electronic states in a 2D Dirac semiconductor reads as 
\begin{eqnarray}\label{hamiltonian}
    H = \begin{pmatrix} 
        \Delta/2 & h_{\bf p} \\ 
        h_{\bf p}^* & -\Delta/2 
    \end{pmatrix},
\end{eqnarray}
where $\Delta$ is a semiconductor band gap and
\begin{align}
h_{\mathbf{p}} &= v p_- + \kappa p_+^2 = v(p_x - i p_y \eta) + \kappa(p_x + i p_y \eta)^2 \nonumber\\
               &= v p e^{-i\eta\varphi_{\mathbf{p}}} + \kappa p^2 e^{2i\eta\varphi_{\mathbf{p}}},
\end{align}
with $\mathbf{p}=(p_x,p_y)$ the particle momentum, $\eta=\pm1$ the valley index, $v$ the linear velocity parameter, $\kappa$ the quadratic-in-momentum coupling coefficient.
Inclusion of both linear ($\propto v$) and quadratic ($\propto\kappa$) in momentum terms in the off-diagonal function $h_p$ is essential to acquire a nontrivial transverse response.
Such a band structure determines the orbital character of the wave functions, which is the fundamental prerequisite for non-zero nonlinear transport.

The corresponding eigenstates and eigenenergies describing the electron states in conduction ($c$) and valence ($v$) bands are
\begin{align}
    \psi_c(\textbf{r}) &= 
    \begin{pmatrix}
        \cos\left(\frac{\theta_{\bf p}}{2}\right) \\
        \sin\left(\frac{\theta_{\bf p}}{2}\right)\frac{h^*_{\textbf{p}}}{|h_\textbf{p}|}
    \end{pmatrix}
\frac{e^{i\textbf{p}\cdot\textbf{r}}}{\sqrt{S}}, \\[6pt]
    \psi_v(\textbf{r}) &= 
    \begin{pmatrix}
        \sin\left(\frac{\theta_{\bf p}}{2}\right) \\
        -\cos\left(\frac{\theta_{\bf p}}{2}\right)\frac{h^*_{\textbf{p}}}{|h_\textbf{p}|}
    \end{pmatrix}
    \frac{e^{i\textbf{p}\cdot\textbf{r}}}{\sqrt{S}},
\end{align}
\begin{gather}
    E_{c,v} = \pm E_{\bf p},\qquad
    E_{\bf p} = \sqrt{\left(\frac{\Delta}{2}\right)^2 + |h_{\bf p}|^2},\\
    |h_{\bf p}|^2 = 
    v^2|\textbf{p}|^2+v\kappa[p_-^3+p_+^3]+\kappa^2|\textbf{p}|^4,
\end{gather}
with $\cos\theta_{\bf p} = \Delta/(2E_{\bf p})$ and $S$ the area of the sample. 
Here, different terms in $|h_{\bf p}|^2$ possess different physical meaning: 
The first term, $v^2|\textbf{p}|^2$, describes the bare isotropic effective mass of electrons' dispersion; the second one, $v\kappa[p_-^3 + p_+^3] = 2v\kappa p^3 \cos(3\varphi_{\bf p})$, acquires the dependence on the angle $3\varphi_\mathbf{p}$, reflecting the trigonal symmetry of the material; and the third, $p^4$ term reflects nonparabolic correction (effective mass renormalization) to the electron dispersion. 
This last term does not play any role in the transport phenomena we study and thus will be dropped from our consideration. 

The term $2v\kappa p^3 \cos(3\varphi_{\bf p})$ produces one of the microscopic contributions to the photocurrent. 
It can be considered weak in comparison with the first term, and thus, the dispersion can be presented in the form $E_{\bf p}\approx\varepsilon_{\bf p}+\delta\varepsilon_{\bf p}$, with $\varepsilon_{\bf p}=p^2/2m\,(m^{-1}=2v^2/\Delta)$ and $\delta\varepsilon_{\bf p}=\eta w(p_x^3-3p_xp_y^2)\,(w=2v\kappa/\Delta)$. 
We call $\delta\varepsilon_{\bf p}$ the warping correction to the electron dispersion, producing the corresponding correction to the electron velocity $\delta{\bf v}_{\bf p}\equiv\nabla_{\bf p}(\delta\varepsilon_{\bf p})$. 

Furthermore, the other microscopic mechanism contributing to the nonlinear photocurrent arises from specific terms in the matrix elements for electrons scattering off impurities,
\begin{equation}\label{Eqnew5}
\begin{split}
    M_{\bf p',p} &= \int d{\bf r}\psi^+_c(\textbf{r})V({\bf r})\psi_c(\textbf{r})
    = V_{\bf p',p}\left[1+i\zeta_{\bf p',p}\right],
\end{split}
\end{equation}
where
\begin{equation}\label{Eqnew6}
\begin{split}
    i\zeta_{\bf p',p} &\approx i\eta\frac{v\kappa}{\Delta^2} \big[ p_y(p_x'^2-p_y'^2) - p_y'(p_x^2-p_y^2) \\
    &\quad + 2p_xp_x'p_y' - 2p_x'p_xp_y \big],
\end{split}
\end{equation}
with $V_{\bf p,p'}$ the Fourier transform of the individual impurity potential and $\zeta_{\mathbf{p}', \mathbf{p}}$ the factor characterizing the scattering asymmetry. 
We distinguish between the theoretical framework of the linear valley Hall effect~\cite{glazov2020skew} and the nonlinear transport discussed in this paper. 
Thus, in what follows, we retain only the terms responsible for the PGE. 
The linear Hall effect is well-described by an isotropic Hamiltonian with only linear in momentum terms, which results in $p^2$-skew scattering amplitude~\cite{glazov2020skew}. 
In contrast, the nonlinear Hall effect requires incorporating $p^2$-terms 
in the Hamiltonian~\eqref{hamiltonian} that capture the trigonal asymmetry. 
These terms lead to $p^3$-skew contributions in the scattering amplitude Eq.~\eqref{Eqnew5}, providing the fundamental skew-scattering mechanism for the nonlinear Hall response. 
To achieve higher precision, we evaluate these scattering processes beyond the Born approximation,
\begin{align}\label{Eqnew7}
    W_{\bf p',p} &= 2\pi N_i\left|M_{\bf p',p}+ \sum_{\bf k}\frac{M_{\bf p',k}M_{\bf k,p}}{\varepsilon_{\bf p}-\varepsilon_{\bf k}+i0}\right|^2 \delta(\varepsilon_{\bf p}-\varepsilon_{\bf p'}) \notag \\
    &\approx 
    W^s_{\bf p',p} + W^a_{\bf p',p}, 
\end{align}
where
\begin{eqnarray}   
    W^s_{\bf p',p} &=& 2\pi N_i\left|V_0\right|^2\delta(\varepsilon_{\bf p}-\varepsilon_{\bf p'}),\\
    \nonumber
    W^a_{\bf p',p} &=& -2\pi N_i\zeta_{\bf p',p}V_0^3 \delta(\varepsilon_{\mathbf{p}} - \varepsilon_{\mathbf{p'}}) \sum_{\bf k}\delta(\varepsilon_{\bf p}-\varepsilon_{\bf k})  
    \\
    &=& -\frac{V_0}{\tau}\zeta_{\bf p',p}\delta(\varepsilon_{\mathbf{p}} - \varepsilon_{\mathbf{p'}}),
\end{eqnarray}
with $\varphi_{\mathbf{p}}$ the polar angle of the momentum,
$N_i$ the concentration of impurities,
$W_{\mathbf{p}', \mathbf{p}}^s$ and $W_{\mathbf{p}', \mathbf{p}}^a$ the symmetric and asymmetric (skew) parts of the scattering probability, respectively,
$V_0$ the strength of the short-range impurity potential, and $\tau=(mN_iV_0^2)^{-1}$ the momentum relaxation time.

It should be noted that both the microscopic mechanisms leading to the transverse current (skew- and warping-related terms) can be assumed small, and thus, we consider them independently, such that they are producing additive terms to the photocurrent.
Hence, considering the anisotropic skew scattering, we will drop out the possible influence of dispersion warping and vice versa.
A further simplification can be made assuming that the isotropic impurity scattering (described by $W^s_{\mathbf{p},\mathbf{p}'}$ probability) can be replaced by a constant $1/\tau$.



\section{Kinetic equations}
The Boltzmann equation governs the kinetics of the system in a static magnetic field $\bf B$ and alternating electric field ${\bf E}(t)$ caused by an external electromagnetic illumination (see details in Supplemental Material~\cite{SMBG}, which includes Refs.~\cite{Xiao2010,Mak2012,Belinicher1980}):
\begin{eqnarray} \label{KineticEq1}
&&\frac{\partial f_{\mathbf{p}}(t)}{\partial t} + e(\mathbf{E}(t) + [\mathbf{V}_{\mathbf{p}} \times \mathbf{B}]) \cdot \nabla_{\mathbf{p}} f_{\mathbf{p}}  \\
\nonumber
&&~~~~~~~~~~~~= -\frac{f_{\mathbf{p}}(t) - n_{\mathbf{p}}}{\tau} + \sum_{\mathbf{p}'} W_{\mathbf{p}', \mathbf{p}}^a f_{\mathbf{p}'}(t),
\end{eqnarray}
with $f_{\mathbf{p}}$ is the nonequilibrium distribution function, 
$e$ the elementary charge,
$\mathbf{E}(t)$ the time-dependent external electric field,
$\mathbf{V}_{\mathbf{p}}=\nabla_{\bf p}E_{\bf p}$ the group velocity of the particles,
and $n_{\mathbf{p}}$ the equilibrium (Fermi-Dirac) distribution function.

Let us first focus on the skew-induced contribution to photocurrents. 
In this case, we assume the electron dispersion and velocities are governed by the bare values $\varepsilon_{\bf p}$ and ${\bf v}_{\bf p}={\bf p}/m$, and we disregard the warping-mediated correction.
Then, Eq.~\eqref{KineticEq1} can be solved in the form of the expansion ${f(\bf p)}={f_1 (\bf p)}+{f_2 (\bf p)}$, with respect to the external electric field strength ${\bf E}(t)$, where ${f_1 (\bf p)}$ is the first-order correction to the equilibrium distribution function, and it obeys the equation
\begin{eqnarray}\label{EqFirst01}
&&\left\{ \frac{i\omega}{\omega_c} + \frac{\partial }{\partial\varphi_{\bf p}} - \frac{1}{\omega_c\tau} \right\} {f_1 (\bf p)}
=\frac{ev}{\omega_c}\left(E_x\cos\varphi_{\bf p}\right.\\
\nonumber
&&\left.~~~~~~~~~~~~~~+E_y\sin\varphi_{\bf p}\right)\frac{\partial n_{\bf p}}{\partial\varepsilon_p} \nonumber
-\frac{1}{\omega_c}\sum_{\bf p'}W^{a}_{\bf p',p}{f_1 (\bf p')},
\end{eqnarray}
and the second-order correction satisfies
\begin{eqnarray}\label{MainEqf2}
&&\left\{ \frac{\partial }{\partial\varphi_{\bf p}} - \frac{1}{\omega_c\tau} \right\} {f_2 (\bf p)} = \frac{e}{\omega_c}{\bf E}\cdot \nabla_{\bf p}{f^*_1 (\bf p)} \nonumber\\
&& ~~~~~~+\frac{e}{\omega_c}{\bf E}^*\cdot \nabla_{\bf p}{f_1 (\bf p)}  
- \frac{1}{\omega_c}\sum_{\bf p'}W^{a}_{\bf p',p}{f_2 (\bf p')}.
\end{eqnarray}

Then, we split the functions $f_{1,2} (\bf p)$ on the symmetric and asymmetric parts~\cite{SMBG}, 
\begin{eqnarray}
f^{(I,II)}_{1,2}({\bf p})=f_{1,2}^{s,(I,II)}({\bf p})+f_{1,2}^{a,(I,II)}({\bf p}).
\end{eqnarray}
The term $f_{2}^{a}$ provides us with the skew scattering-mediated longitudinal and transverse (Hall) electric current densities.

Furthermore, let us switch to the warping-mediated photocurrents. 
In this case, we drop the anisotropic contribution to electron scattering off impurities, account for warping-induced corrections to electron energy and velocity, and employ a perturbative analysis of the Boltzmann equation.
Here, for distinguishability, all the quantities will be marked by the subscript $w$. 
The Boltzmann kinetic equations governing the first- and second-order corrections to the distribution function, $f_{1w}$ and $f_{2w}$, with respect to the external electric field read as
\begin{gather}
\frac{\partial f_{1w}}{\partial t} + e [\mathbf{V}_{\mathbf{p}} \times \mathbf{B}] \cdot \nabla_{\mathbf{p}} f_{1w} + \frac{f_{1w}}{\tau}
\\\nonumber
=-e \mathbf{E}(t) \cdot \nabla_{\mathbf{p}} n(E_{\mathbf{p}})
\end{gather}
and
\begin{gather}
\frac{\partial f_{2w}}{\partial t} + e [\mathbf{V}_{\mathbf{p}} \times \mathbf{B}] \cdot \nabla_{\mathbf{p}} f_{2w} + \frac{f_{2w}}{\tau}\\ \nonumber
=-e \mathbf{E}(t) \cdot \nabla_{\mathbf{p}} f_{1w}.
\end{gather}
Furthermore, we separate the distribution function into isotropic and angle-dependent warping-induced corrections: $f_{1w}=f^0_{1w}+\delta f_{1w},\,f_{2w}=f^0_{2w}+\delta f_{2w}$. 
Here, $f_{2w}^0$ satisfies
\begin{equation}
\left( \frac{1}{\tau} + e [\mathbf{v}_{\bf p} \times \mathbf{B}] \cdot \nabla_{\mathbf{p}} \right) f_{2w}^0 = -2e \operatorname{Re} \left\{ \mathbf{E}^* \cdot \nabla_{\mathbf{p}} f_{1w}^0 \right\},
\end{equation}
whereas $\delta f_{2w}$ is determined by
\begin{eqnarray}
&&\left(\frac{1}{\tau} - \omega_c \frac{\partial}{\partial\varphi}\right) \delta f_{2w}
=-2e \operatorname{Re}\left\{ \mathbf{E}^* \cdot \nabla_{\mathbf{p}} \delta f_{1w} \right\}\nonumber\\
&&~~~~~~~~~~~~~~~~-e[\delta\mathbf{v}_{\mathbf{p}} \times \mathbf{B}] \cdot \nabla_{\mathbf{p}} f_{2w}^0,
\end{eqnarray}
where the first term on the right-hand side accounts for the action of the electric field on the linear warping correction $\delta f_{1w}$, and the second term describes the Lorentz force contribution arising from the anomalous velocity $\delta \mathbf{v}_p$.
The corresponding expressions for $f_{1w}^0$ and $\delta f_{1w}$ are anaylized in~\cite{SMBG}.


\section{Electric current}
The analytical expressions for skew-induced components of the electric current density can be derived from the anisotropic corrections to the first- and second-order distribution functions, yielding:
\begin{equation}
{\bf j}^{\eta}_s = e\int \mathbf{v_p} f_{2}^{a,(I,II)}(\mathbf{p}) \frac{d^2\mathbf{p}}{(2\pi\hbar)^2}.
\end{equation}
This provides two contributions within the second-order correction $f_2^{a(I,II)}({\bf p})$: one stemming from the driving term $\mathbf{E} \cdot \nabla_{\mathbf{p}} f_1(\mathbf{p})$ involving the first-order correction (let us call it $Is$), and the other ($IIs$) originating from the asymmetric scattering term in \eqref{MainEqf2}.


Then, $(Is)$ contribution yields the transport coefficients (see~\eqref{maineq}):

\begin{equation}
\label{EqContrib1x}
\begin{split}
\chi_{\eta}^{(Is)} = &\frac{\mathcal{J}_s}{(1+\omega_c^2\tau^2)(1+4\omega_c^2\tau^2)} \\
&\times \frac{1 + \tau^2(\omega^2 + 2\omega_c^2) + \tau^4\omega_c^2(3\omega^2 + \omega_c^2)}{[1 + \tau^2(\omega_c^2 - \omega^2)]^2 + 4\omega^2\tau^2},
\end{split}
\end{equation}

\begin{equation}
\label{EqContrib1y}
\begin{split}
\lambda_{\eta}^{(Is)} = &-\frac{2\omega_c \tau \mathcal{J}_s}{(1+\omega_c^2\tau^2)(1+4\omega_c^2\tau^2)} \\
&\times \frac{1 + 2\omega_c^2\tau^2 + \tau^4\omega_c^2(\omega_c^2 - \omega^2)}{[1 + \tau^2(\omega_c^2 - \omega^2)]^2 + 4\omega^2\tau^2},
\end{split}
\end{equation}
where
%


\begin{equation}
    \mathcal{J}_s = -\eta \frac{7 e^3 p_F^5 V_0 w \tau^2}{8\sqrt{2} \pi^2 \hbar^4 \sqrt{m} \Delta^{3/2}}.
\end{equation}

\begin{figure*}[t]
    \centering
\includegraphics[width=\textwidth,height=8cm, keepaspectratio]{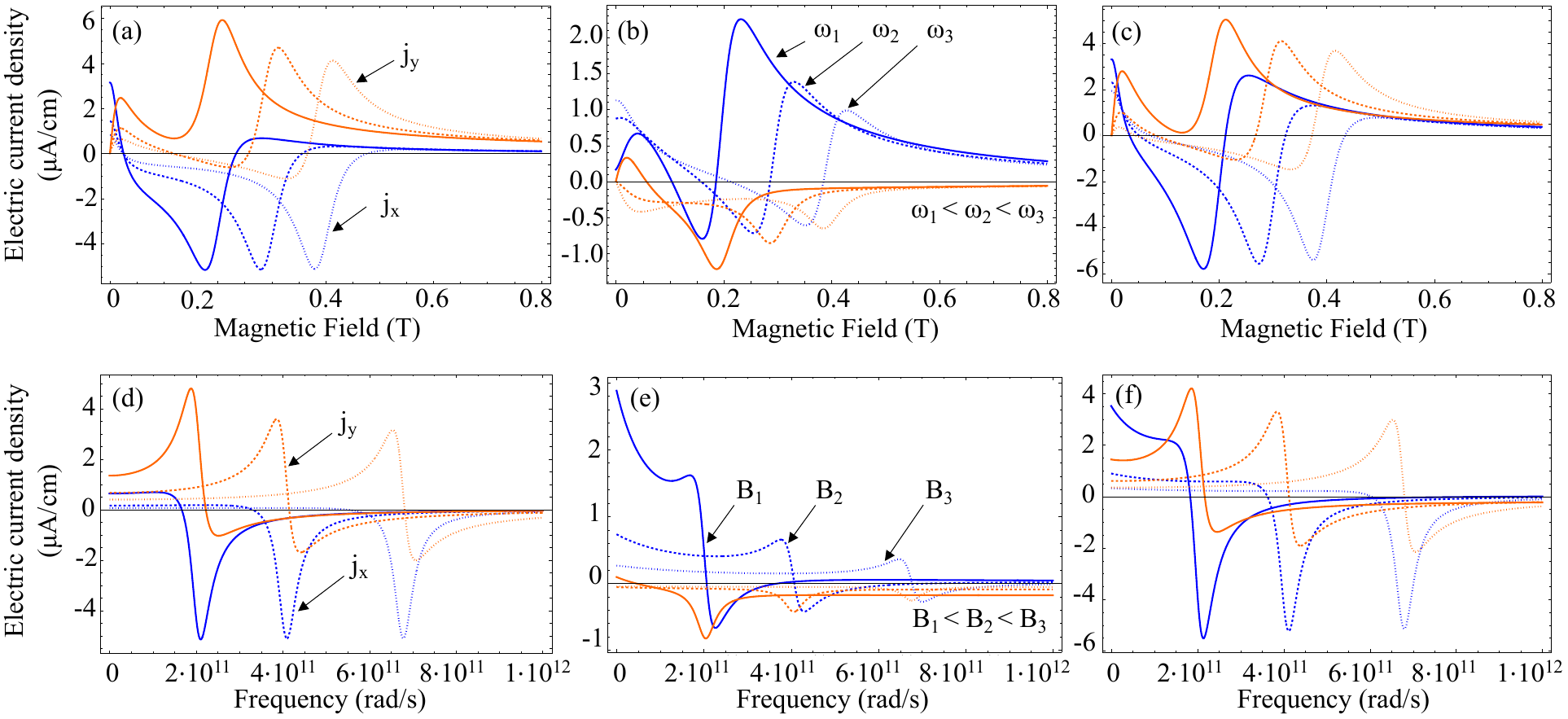}
\captionsetup{justification=raggedright,singlelinecheck=false}
\caption{
    The cyclotron resonance mediated by the skew scattering and warping mechanisms: Electric current density as a function of the external magnetic field (a--c) and external light frequency (d--f) at $\theta=0$. 
Blue and orange curves represent the longitudinal ($j_x$) and transverse ($j_y$) components, respectively. 
In (a--c), solid, dashed, and dotted lines correspond to $\omega_1\tau =5$, $\omega_2\tau=7.5$ and $\omega_3\tau=10$; 
in (d--f), the same styles denote magnetic field strengths 
$B_1 = 0.3~\text{T}$, 
$B_2 = 0.6~\text{T}$ and 
$B_3 = 1.0~\text{T}$.
Panels (a) and (d) correspond to the first contribution $j^{\eta(I)}=j^{\eta(Is)}+j^{\eta(Iw)}$, panels (b) and (e) correspond to the second contribution $j^{\eta(II)}=j^{\eta(IIs)}+j^{\eta(IIw)}$, and panels (c) and (f) correspond to the total current density. Data used for plotting the graphs are presented in Table \ref{tab:parameters}.}
    \label{fig:resonance}
\end{figure*}

The $(IIs)$ contribution reads as
\begin{equation}
\label{EqContrib2x}
\chi^{(IIs)}_\eta = \frac{2\tilde{\mathcal{J}}_s \left[ (1 + \tau^2(\omega_c^2 - \omega^2))^2 - 4 \omega^2 \tau^2 \right]}{(1 + \omega_c^2\tau^2) \left( [1 + \tau^2(\omega_c^2 - \omega^2)]^2 + 4\omega^2\tau^2 \right)},
\end{equation}
\begin{equation}
\label{EqContrib2y}
\lambda^{(IIs)}_\eta = - \omega_c \tau \frac{2\tilde{\mathcal{J}}_s \left[ (1 + \tau^2(\omega_c^2 - \omega^2))^2 - 4 \omega^2 \tau^2 \right]}{(1 + \omega_c^2\tau^2) \left( [1 + \tau^2(\omega_c^2 - \omega^2)]^2 + 4\omega^2\tau^2 \right)},
\end{equation}
%
%
%
%
where $\tilde{\cal J}_s=-{\cal J}_s/7$.

The electric current density generated by the trigonal warping, ${\bf j}_w$, also consists of two contributions:
\begin{equation}
    \mathbf{j}^{\eta}_{w} = e \sum_{\mathbf{p}} \mathbf{V}_{\mathbf{p}} f_{2w} \approx e \sum_{\mathbf{p}} \delta\mathbf{v}_{\mathbf{p}} f_{2w}^0 + e \sum_{\mathbf{p}} \mathbf{v}_{\mathbf{p}} \delta f_{2w}.
\end{equation}
We can also derive the warping transport coefficients similar to~\eqref{maineq}~\cite{SMBG}, yielding two terms ($Iw$) and ($IIw$), respectively:
\begin{align}
\label{eqchi-1w}
\chi^{(Iw)}_\eta &= \frac{3 {\cal J}_w \left[ 1 + \omega^2\tau^2 + 3\omega_c^2\tau^2 + 2\omega_c^2\tau^2(\omega_c^2\tau^2 - \omega^2\tau^2) \right]}{(1 + 4\omega_c^2\tau^2) \left[ (1 - \omega^2\tau^2 + \omega_c^2\tau^2)^2 + 4\omega^2\tau^2 \right]},
\end{align}
\begin{align}
\label{eqla-1w}
\lambda^{(Iw)}_\eta &= \frac{3 {\cal J}_w \, \omega_c\tau \left[ 1 + 3\omega^2\tau^2 + \omega_c^2\tau^2 \right]}{(1 + 4\omega_c^2\tau^2) \left[ (1 - \omega^2\tau^2 + \omega_c^2\tau^2)^2 + 4\omega^2\tau^2 \right]},
\end{align}
where
\begin{equation}
{\cal J}_w = \eta\frac{e^3 w \tau^2 p_F^2}{4\pi \hbar^2}.
\end{equation}

Similarly, 
\begin{eqnarray}
\label{eqchi-2w}
&&~~~~~~\chi^{(IIw)}_\eta= -\frac{3 {\cal J}_w}{2 \left[1 + (2\omega + 2\omega_c)^2\tau^2\right]} \times 
\\
&&\left( \frac{6\omega_c^2\tau^2 \left( 1 - \omega^2\tau^2 + \omega_c^2\tau^2 \right)}{\left[1 + (\omega - \omega_c)^2\tau^2\right] \left[1 + (\omega + \omega_c)^2\tau^2\right]} - \frac{1 - 2\omega_c^2\tau^2}{1 + \omega_c^2\tau^2} \right),\nonumber
\end{eqnarray}
\begin{eqnarray}
\label{eqla-2w}
&&~~~~~~\lambda^{(IIw)}_\eta= \frac{9 {\cal J}_w \, \omega_c\tau}{2 \left[1 + (2\omega + 2\omega_c)^2\tau^2\right]} \times  \\
&& \left( \frac{1}{1 + \omega_c^2\tau^2} + \frac{1 - 3\omega^2\tau^2 + \omega_c^2\tau^2}{\left[1 + (\omega - \omega_c)^2\tau^2\right] \left[1 + (\omega + \omega_c)^2\tau^2\right]} \right).\nonumber
\end{eqnarray}

The total valley electric current density can be found as the joint contribution: $\mathbf{j}^\eta=\mathbf{j}_s^{\eta}+\mathbf{j}_w^{\eta}$.
In addition to the valley-polarized photocurrent, we can evaluate the total photocurrent, defined as the sum of contributions from both valleys: $\mathbf{j} = \sum_{\eta = \pm 1} \mathbf{j}^{\eta}$.


\section{Results and discussion}
Let us analyze how the system's symmetry dictates the form of the photoresponse.
If $B=0$, the cyclotron frequency $\omega_c$ vanishes, and Eqs.~\eqref{EqContrib1x}--\eqref{EqContrib2y},~\eqref{eqchi-1w}--\eqref{eqla-2w} reduce to:
\begin{eqnarray}
&&j^{\eta}_x= \left[ \frac{\mathcal{J}_s(1+\omega^2\tau^2) - 8\tilde{\mathcal{J}}_s\omega^2\tau^2}{(1+\omega^2\tau^2)^2+4\omega^2\tau^2} \right. \nonumber\\
&& \left. + \frac{3 \mathcal{J}_w (1 + 7\omega^2\tau^2)}{2 (1 + \omega^2\tau^2)(1 + 4\omega^2\tau^2)} \right] (|E_x|^2 - |E_y|^2)
\end{eqnarray}
\begin{eqnarray}
&&j^{\eta}_y= - \bigg[ \frac{\mathcal{J}_s(1+\omega^2\tau^2) - 8\tilde{\mathcal{J}}_s\omega^2\tau^2}{(1+\omega^2\tau^2)^2+4\omega^2\tau^2} \nonumber\\
&& + \frac{3 \mathcal{J}_w (1 + 7\omega^2\tau^2)}{2 (1 + \omega^2\tau^2)(1 + 4\omega^2\tau^2)} \bigg] (E_x E_y^* + E_y E_x^*),
\end{eqnarray}
so, $j^{\eta}_x=\chi_\eta(0)(|E_x|^2 - |E_y|^2)$
and $j^{\eta}_y=-\chi_\eta(0)(E_x E_y^* + E_y E_x^*)$, and $\lambda_\eta(0)=0$.
This structure is a direct reflection of the $C_{3h}$ symmetry of the individual valley~\cite{liu2023linear, kovalev2024role}.

%

%
%

%
%
%
\begin{figure*}[t]
    \centering
\includegraphics[width=\textwidth, height=8cm, keepaspectratio]{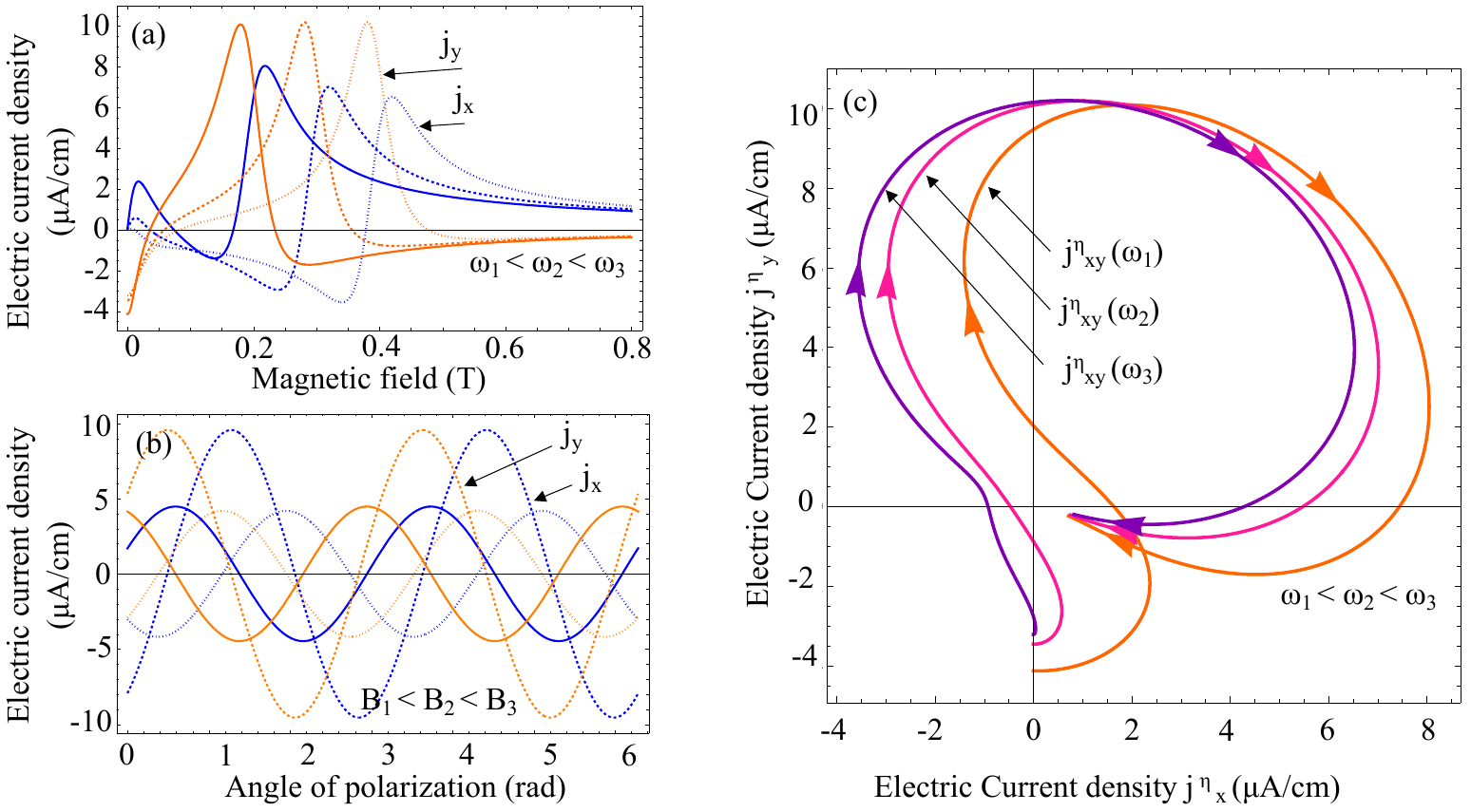}
\captionsetup{justification=raggedright,singlelinecheck=false}
\caption{Photocurrent dependence on the external light polarization.
(a) The components of the valley electric current density as functions of external magnetic field for the polarization $\theta=$ 45 degrees, $j^{\eta}_x$ (blue) and $j^{\eta}_y$ (orange) for $\omega\tau$ = 5 (solid), 7.5 (dashed), and 10 (dotted). 
(b) Dependence of $j_x$ and $j_y$ on the degree of polarization $\theta$.
\\
(c) Hodograph of the current density vector $\mathbf{j}^{\eta}$ for valley currents, illustrating its trajectory and the phase interplay between the longitudinal and transverse components. Data used for plotting the graphs are presented in Table \ref{tab:parameters}.}
\label{fig:pol}
\end{figure*}

Figure~\ref{fig:resonance} shows that the valley-selective response is highly sensitive to external fields (by numerical calculations). 
\begin{table}[!b]
\centering
\caption{Parameters used in the calculations.}
\label{tab:parameters}
\begin{tabular}{@{} l l l @{}}
\toprule
Parameter & Symbol & Value (SI units) \\
Effective mass            & $m$      & $0.26\cdot m_e$\\
Fermi momentum            & $p_F$    & $8.4\times10^{-26}$ kg\,m/s\\
Band gap     & $\Delta$ & $3.2\times10^{-19}$ J\\
Relaxation time           & $\tau$   & $3.8\times10^{-11}$ s \\
Carrier concentration     & $n$      & $10^{13}$ cm$^{-2}$\\
Impurity potential strength & $V_0$ & $10^{-37}$ $\text{ J} \cdot \text{m}^2$ \\
Electric field intensity & $E_0$ & $10^{2}-10^{3}$~V/m\\
\end{tabular}
\end{table}
In particular, the dependence of the current on the magnetic field for fixed frequencies [Fig.~\ref{fig:resonance}(a--c)] shows a clear sign-changing behavior: at low fields, the current is negative, but as the field increases, it rapidly changes sign and forms a sharp resonant peak. 
These peaks reach magnitudes of around $5$~$\mu A/cm$, which is significantly higher than the current far from the resonance, suggesting a highly efficient conversion of light energy into the current, particularly the nonlinear valley Hall current. 
These resonance peaks shift toward higher magnetic fields as the excitation frequency $\omega$ increases [see solid to dotted lines in Fig.~\ref{fig:resonance}(a)]. This shift follows the cyclotron resonance condition, $\omega \approx \omega_c$.

Comparing the two main contributions ($I$ and $II$) to electric current (including both the $s$ and $w$ mechanisms), we see that $j^{\eta(I)}$ determines the overall scale of the effect, while $j^{\eta(II)}$ [Fig.~\ref{fig:resonance}(b)] is smaller in magnitude.
Similar behavior takes place in the spectrum of the current density for fixed cyclotron frequencies (magnetic fields) [Fig.~\ref{fig:resonance}(d--f)].



Table~\ref{tab:materials_comparison} shows the behavior of the trigonal warping-mediated ($j^{(w)}$) and skew scattering-mediated ($j^{(s)}$) currents for $\text{WS}_2$ as the carrier density $n$ increases, evaluated at a light frequency of $\omega = 1.3 \times 10^{11}\,\text{rad/s}$. 
In the low-density regime ($n = 10^{11}\,\text{cm}^{-2}$), the photocurrent is strongly dominated by trigonal warping ($j^{(w)} \gg j^{(s)}$). 
Specifically, $j^{(w)}$ is approximately 3 orders of magnitude larger than $j^{(s)}$, with the ratio $\vert{}j^{(s)}/j^{(w)}\vert{}$ on the order of $\sim 10^{-3}$. 
Indeed, at low carrier densities, the Fermi surface is small, which strongly suppresses asymmetric impurity scattering.

\begin{table}[b!] 
\centering
\caption{Comparison of current components $j_x$ and $j_y$ for $\text{WS}_2$ under various doping levels.}
\label{tab:materials_comparison}
\footnotesize
\setlength{\tabcolsep}{1pt}
\begin{tabular}{llcccc}
\toprule
\textbf{Material} & \textbf{} & $n$ & $j^{(s)}$ & $j^{(w)}$ & \textbf{Ratio} \\
& & ($\text{cm}^{-2}$) & ($\text{$\mu$A/cm}$) & ($\text{$\mu$A/cm}$) & $|j^{(s)}/j^{(w)}|$ \\
\midrule
\multirow{6}{*}{$\text{WS}_2$} 
 & $j_x$ & \multirow{2}{*}{$10^{11}$} & $-1.0 \cdot 10^{-4}$ & $+4\cdot10^{-2}$ & \multirow{2}{*}{$\sim 10^{-3}$} \\
 & $j_y$ & & $+6.4 \cdot 10^{-5}$ & $+3\cdot10^{-2}$ & \\
\cmidrule{2-6}
 & $j_x$ & \multirow{2}{*}{$10^{12}$} & $-2.0\cdot10^{-2}$ & $3.82\cdot10^{-1}$ & \multirow{2}{*}{$\sim 10^{-1}$} \\
 & $j_y$ & & $+1.28\cdot10^{-2}$ & $+2.71\cdot10^{-1}$ & \\
\cmidrule{2-6}
& $j_x$ & \multirow{2}{*}{$10^{13}$} & $-10.1$ & $3.99$ & \multirow{2}{*}{$\sim 10^{0}$} \\
& $j_y$ & & $+6.4$ & $+2.83$ & \\
\bottomrule
\end{tabular}
\end{table}
Increasing the carrier density (for example, using a single-gate FET allowing for $n = 10^{12}\,\text{cm}^{-2}$ or a dual-gate h-BN structure, where $n = 10^{13}\,\text{cm}^{-2}$) enhances $j^{(s)}$ by several orders of magnitude. Consequently, the ratio $\vert{}j^{(s)}/j^{(w)}\vert{}$ increases from $\sim 10^{-3}$ at $n = 10^{11}\,\text{cm}^{-2}$ up to $\sim 10^{0}$ (becoming comparable in magnitude) at $n = 10^{13}\,\text{cm}^{-2}$. Although trigonal warping remains dominant or comparable across the considered range, the skew-scattering contribution becomes significantly more pronounced at higher doping levels.


Crucially, at $n = 10^{13}\,\text{cm}^{-2}$, the Fermi level in $\text{WS}_2$ remains below the bottom of the side Q-valleys ($E_F < \Delta_\textrm{KQ}$). 
This confirms that the system stays in the single-valley (K/K') regime and validates our two-band kinetic model.


Figure~\ref{fig:pol} demonstrates how light polarization affects electric current. 
For the $\theta=$ 45-degree polarization, the behavior of the current with magnetic field changes as compared to the results in Fig.~\ref{fig:resonance}(c), in particular, the regions where the sign of the electric current changes are different (Fig.~\ref{fig:pol}(a)). 
The valley current from both mechanisms reaches its maximum of about 10~$\mu$A/cm at this polarization.
Figure~\ref{fig:pol}(b) show that both $j^{\eta}_x$ and $j^{\eta}_y$ experience oscillations as functions of the polarization angle $\theta$. 
To better understand this interplay, we built the current vector hodographs [Fig.~\ref{fig:pol}(c),(d)]. 
This plot acts as a map, tracing the tip of the total current vector $\mathbf{j} = (j_x, j_y)$. 
The twisted trajectory of the curve is a classic signature of cyclotron resonance. 
When the magnetic field is far from the resonant value, the current stays small, and the curve remains close to the origin. 
As the system approaches the resonance condition ($  \omega \approx \omega_c  $), the current vector grows sharply and rotates, forming closed loops. 
These loops highlight the change in the phase relationship between the longitudinal and transverse components in the strong-resonance regime.

The shape and position of each loop reveal important features of the response. For example, when a loop is elongated along the $  j_y  $ axis, the Hall (transverse) current dominates over the longitudinal one at specific field values. Crossings of the $  j_x=0  $ axis mark points where the longitudinal current vanishes completely, leaving only a pure transverse (Hall) response.
Finally, the transitions between quadrants (from positive to negative $  j_y  $) demonstrate the sign reversal of the transverse current, which can occur across resonance or due to valley switching $\eta\rightarrow-\eta$ in the 2D semiconductor.

It should be noted that the magnetic field induces valley polarization via Zeeman and spin-orbit splittings of the electronic spectrum within each valley. 
Since the external alternating electric field \(\mathbf{E}\) does not induce spin-flip processes, it leaves the cyclotron resonance within a given valley unaffected. 
Consequently, our analysis holds for electrons with a fixed spin polarization. 
Although the magnetic field-induced shift of spin-split states can redistribute electrons among these states within each valley, thereby generating a finite net current, focusing exclusively on the valley cyclotron resonance allows us to disregard the spin-dependent dynamics of the valley electrons.

Beyond the spin-induced valley energy shifts, another contribution arises from the orbital magnetic moment of carriers in each valley~\cite{PhysRevB.107.235417}.
This magnetic moment induces a Zeeman-like shift of the energy levels, which takes opposite signs in different valleys and equally displaces all states. 
While this shift does not noticeably alter the valley current or the cyclotron resonance, its valley-asymmetric nature prevents the mutual cancellation of valley currents (since it can be absorbed into the valley-specific Fermi energy), thereby generating a net total current in the system.

In the absence of a magnetic field, time-reversal symmetry enforces an exact cancellation between the two valleys, leading to a zero net charge current. 
However, an out-of-plane magnetic field $B_z$ couples to the carrier orbital magnetic moment $m_z$, shifting the Fermi level in opposite directions in the two valleys: $E_F^\eta = E_F + \eta m_z B_z$.
Expanding the response coefficients to linear order in $m_z B_z / E_F \ll 1$, we obtain the non-zero total charge photocurrent coefficients:
\begin{equation}
\chi = \frac{m_z B_z}{E_F} \left[ 5 \chi_{\eta=1}^{(s)} + 2 \chi_{\eta=1}^{(w)} \right],
\end{equation}
\begin{equation}
\lambda = \frac{m_z B_z}{E_F} \left[ 5 \lambda_{\eta=1}^{(s)} + 2 \lambda_{\eta=1}^{(w)} \right].
\end{equation}
Here, $\chi_{\eta=1}^{(s,w)}$ and $\lambda_{\eta=1}^{(s,w)}$ denote the contributions from skew scattering ($s$) and trigonal warping ($w$) in the $\eta = +1$ valley.
Due to the small scale factor $m_z B_z / E_F \sim 10^{-3}$ at $B_z = 0.3\text{ T}$, the total charge photocurrent is scaled down compared to the valley photocurrent ($j^{\eta} \approx 5\text{$\mu$A/cm}$), yielding a net macroscopic photocurrent of $j \approx 0.03\text{ $\mu$A/cm}$. This macroscopic current can be directly measured in standard photogalvanic experiments.

%
%
%



\section{Conclusions}
We developed a kinetic theory to describe the nonlinear (anomalous) longitudinal and Hall photoresponses in Dirac materials exposed to an EM field of external light and an alternating permanent magnetic field, accounting for the extrinsic (skew) scattering of electrons on impurities and the trigonal warping of the valleys. 
For that, we employed an effective two-band Hamiltonian incorporating both linear- and quadratic-in-momentum terms, which are essential for breaking the inversion symmetry and accurately reflecting the $C_{3h}$ trigonal symmetry of the valley, and evaluated the scattering matrix elements.
Then, we used the semiclassical Boltzmann transport equations to compute corrections to the nonequilibrium particle distribution function and subsequently evaluated the electric current density, examining its dependence on the external-field frequency, polarization, and magnetic field. 

The modeling demonstrated the emergence of strong cyclotron resonance peaks when the excitation frequency matches the cyclotron frequency. 
Collectively, these findings demonstrate precise control of phase-sensitive transport via magnetic field modulation and spectral tuning, offering a framework for high-frequency magneto-electronic devices. 
Our theory captures the cyclotron resonance, revealing a strong resonant enhancement of the nonlinear valley Hall current that is absent in lower orders.
The predicted resonances occur in experimentally accessible magnetic fields and THz frequencies, making them readily verifiable in state-of-the-art TMD monolayers and van der Waals heterostructures.


\section*{Acknowledgements}
We were supported by the National Natural Science Foundation of China (NSFC) under Grant No.~W2532001, Guangdong Basic and Applied Basic Research Foundation under Grant No.~2026A1515012415,
the Ministry of Science and Higher Education of the Russian Federation (Project FSUN-2026-0004), and the Foundation for the Advancement of Theoretical Physics and Mathematics ``BASIS''.
The authors thank Elizaveta Osipova for the help with the figures.

\bibliography{bib}
\bibliographystyle{apsrev4-2}

\end{document}


\title{Supplemental Material for ``Resonant nonlinear THz valley Hall effect\\ in two-dimensional Dirac semiconductors''}


\author{V. N.~Ivanova}
\affiliation{Technion -- Israel Institute of Technology, Haifa, 3200003, Israel}
\affiliation{Guangdong Technion-Israel Institute of Technology, 241 Daxue Road, Shantou, Guangdong, China, 515063}


\author{V.~M.~Kovalev}
\affiliation{Novosibirsk State Technical University, Novosibirsk 630073, Russia}

\author{O.~V.~Kibis}
\affiliation{Novosibirsk State Technical University, Novosibirsk 630073, Russia}

\author{I.~G.~Savenko}
\affiliation{Guangdong Technion-Israel Institute of Technology, 241 Daxue Road, Shantou, Guangdong, China, 515063}
\affiliation{Technion -- Israel Institute of Technology, Haifa, 3200003, Israel}

\date{\today}

\begin{abstract}
In this Supplemental Material, we provide a detailed theoretical framework for calculating the resonant nonlinear valley Hall effect in 2D Dirac semiconductors. 
We start with the derivation of the two-band effective Hamiltonian and the corresponding scattering matrix elements, accounting for trigonal warping and inversion-symmetry breaking. 
Furthermore, we present a step-by-step solution of the Boltzmann kinetic equation in the presence of a magnetic field and THz radiation. 
Special attention is paid to the roles of skew-scattering and trigonal warping mechanisms, and to the derivation of the second-order asymmetric distribution function, which governs the giant nonlinear current response near the cyclotron resonance.
\end{abstract}

\maketitle

\tableofcontents

\section{Effective Hamiltonian, eigenstates, and scattering matrix elements}
We consider the Hamiltonian
\begin{gather}
H=
\left(\begin{array}{cc}
\Delta/2                  &               h_{\bf p} \\
h_{\bf p}^*       &                -\Delta/2
\end{array}\right)
\textrm{, where}\\
\nonumber
h_{\bf p}=vp_-+\kappa p_+^2=v(p_x-ip_y\eta)+\kappa(p_x+ip_y\eta)^2=vpe^{-i\eta\varphi_{\bf p}}+\kappa p^2 e^{2i\eta\varphi_{\bf p}}.
\end{gather}
The Hamiltonian parameters $v$ and $\kappa$ are chosen to reflect the electronic structure of MoS$_2$ monolayers \cite{Mak2012}, and the eigenstates are defined following the standard formalism for Berry phase effects in 2D systems \cite{Xiao2010}.
The eigenstates and eigenenergies are
\begin{gather}
\psi_c(\textbf{r})=\left(
                     \begin{array}{c}
                       \cos\left(\frac{\theta_{\bf p}}{2}\right) \\
                       \sin\left(\frac{\theta_{\bf p}}{2}\right)\frac{h^*_{\textbf{p}}}{|h_\textbf{p}|} \\
                     \end{array}
                   \right)\frac{e^{i\textbf{pr}}}{\sqrt{S}},\,\,\,\,\,\,\,\,\,\,\,
\psi_v(\textbf{r})=\left(
                     \begin{array}{c}
                       \sin\left(\frac{\theta_{\bf p}}{2}\right) \\
                       -\cos\left(\frac{\theta_{\bf p}}{2}\right)\frac{h^*_{\textbf{p}}}{|h_\textbf{p}|} \\
                     \end{array}
                   \right)\frac{e^{i\textbf{pr}}}{\sqrt{S}},\\\nonumber
E_{c,v} = 
    \pm E_{\bf p},\,\,\,\,\,\,\,
    E_{\bf p}=
    \sqrt{\left(\frac{\Delta}{2}\right)^2+|h_{\bf p}|^2},
    \,\,\,\,\,\,\,\,\,\,
    |h_{\bf p}|^2 = 
    v^2|\textbf{p}|^2+v\kappa
    [p_-^3+p_+^3]+\kappa^2|\textbf{p}|^4,
\end{gather}
where $\cos\theta_{\bf p}=\Delta/2E_{\bf p}$, $S$ is a sample area, and 
$v\kappa[p_-^3+p_+^3]=2v\kappa p^3\cos(3\varphi_{\bf p})$.

The scattering matrix element is
\begin{gather}\label{Eqnew5}
    M_{\bf p',p}=\int d{\bf r}\psi^+_c(\textbf{r})V({\bf r})\psi_c(\textbf{r})=
    V_{\bf p',p}\left[
    \cos\left(\frac{\theta_{\bf p'}}{2}\right) \cos\left(\frac{\theta_{\bf p}}{2}\right)+
     \sin\left(\frac{\theta_{\bf p'}}{2}\right) \sin\left(\frac{\theta_{\bf p}}{2}\right)
     \frac{h_{\bf p'}}{|h_{\bf p'}|}\frac{h^*_{\textbf{p}}}{|h_\textbf{p}|}
     \right]\\\nonumber
    =V_{\bf p',p}\left[1+\frac{h_{\bf p'}h^*_{\bf p}}{\Delta^2}\right]
    =V_{\bf p',p}\left[1+\frac{1}{\Delta^2}
    \left(vpe^{-i\eta\varphi_{\bf p}}+\kappa p^2 e^{2i\eta\varphi_{\bf p}}\right)
    \left(vp'e^{i\eta\varphi_{\bf p'}}+\kappa p'^2 e^{-2i\eta\varphi_{\bf p'}}\right)
    \right]=V_{\bf p',p}\left[1+\Lambda_{\bf p',p}\right],
\end{gather}
where
\begin{gather}
\label{Eqnew6}
   \Lambda_{\bf p',p}=\frac{v^2}{\Delta^2}[p_x'p_x+p_y'p_y+{i\eta(p_x'p_y-p_y'p_x)}]\\\nonumber
   +\frac{\mu v}{\Delta^2}[p_x'(p_x^2-p_y^2)+p_x(p_x'^2-p_y'^2)-2p_y'p_xp_y-2p_yp_x'p_y'
   +{i\eta[p_y(p_x'^2-p_y'^2)-p_y'(p_x^2-p_y^2)+2p_xp_x'p_y'-2p_x'p_xp_y]}]\\
   \nonumber
   \approx{i\eta\frac{\kappa v}{\Delta^2}[p_y(p_x'^2-p_y'^2)-p_y'(p_x^2-p_y^2)+2p_xp_x'p_y'-2p_x'p_xp_y]}=i\zeta_{\bf p',p}.
\end{gather}
~
~
\\
Thus, the amplitude of electron-impurity scattering can be written in a simple form: 
$ M_{\bf p',p}=V_{\bf p',p}\left[1+i\zeta_{\bf p',p}\right]$. 
Here, we retain only terms responsible for the PGE effect. 
The total scattering probability reads as
\begin{gather}
\label{Eqnew7}
   W_{\bf p',p}=2\pi N_i\left|M_{\bf p',p}+
   \sum_{\bf k}\frac{M_{\bf p',k}M_{\bf k,p}}{\varepsilon_{\bf p}-\varepsilon_{\bf k}+i\delta}\right|^2
   \delta(\varepsilon_{\bf p}-\varepsilon_{\bf p'})\approx W^s_{\bf p',p}+W^a_{\bf p',p},~\textrm{where}\\
   W^s_{\bf p',p}=2\pi N_i\left|V_0\right|^2\delta(\varepsilon_{\bf p}-\varepsilon_{\bf p'}),
   \end{gather}
   \begin{gather}
   \label{EqAprob}
   W^a_{\bf p',p}=-2\pi N_i\zeta_{\bf p',p}V_0^3
\delta(\varepsilon_{\mathbf{p}} - \varepsilon_{\mathbf{p'}})
   \sum_{\bf k}\delta(\varepsilon_{\bf p}-\varepsilon_{\bf k})
   =-m N_iV_0^3\zeta_{\bf p',p}\delta(\varepsilon_{\mathbf{p}} - \varepsilon_{\mathbf{p'}})=-\frac{V_0}{\tau}\zeta_{\bf p',p}\delta(\varepsilon_{\mathbf{p}} - \varepsilon_{\mathbf{p'}}).
\end{gather}
The last two expressions are for the case of the short-range impurities:  $V_{\bf p',p}=V_0$.

\section{Skew-scattering mediated kinetic equations}
The equation, which governs the kinetics of the system, reads as
\begin{gather}\label{KineticEq1}
  \frac{\partial f_{\bf p}(t)}{\partial t}
  +{e({\bf E}(t)+[{\bf v}_{\bf p}\times{\bf B}])
\cdot\nabla_{\bf p}f_{\bf p}
}
  =-\frac{f_{\bf p}(t)-n_{\bf p}}{\tau}+\sum_{\bf p'}W^{a}_{\bf p',p}f_{\bf p'}(t),
\end{gather}
where the relation $f_{\bf p}\sum_{\bf p'}W^a_{\bf p',p}=0$ was assumed, and
\begin{eqnarray}
{\bf E}(t)={\bf E}e^{-i\omega t}+{\bf E}^*e^{i\omega t},\\
\nabla_{\bf p}f_{\bf p}=\frac{\partial f}{\partial p}\mathbf{e}_{\bf p}+\frac{1}{p}\frac{\partial f}{\partial\varphi_{\bf p}}\mathbf{e}_{\varphi_{\bf p}}.
\end{eqnarray}
Therefore, the Lorentz force term in the kinetic equation becomes:
\[
e[\mathbf{v}_{\mathbf{p}} \times \mathbf{B}] \cdot \nabla_{\mathbf{p}} f_{\mathbf{p}} = -\omega_c \frac{\partial f_{\mathbf{p}}}{\partial \varphi}.
\]
This derivation shows how the 2D vector cross-product reduces to a simple angular derivative in the polar coordinates:
\begin{eqnarray}
\nonumber
\mathbf{e}_\mathbf{p}&=&\cos\varphi_\mathbf{p}\mathbf{e}_\mathbf{x}+\sin\varphi_\mathbf{p}\mathbf{e}_\mathbf{y},\\
\nonumber
\mathbf{e}_{\varphi_\mathbf{p}}&=&-\sin\varphi_\mathbf{p}\mathbf{e}_\mathbf{x}+\cos\varphi_\mathbf{p}\mathbf{e}_\mathbf{y}.
\end{eqnarray}

Furthermore, we expand the nonequilibrium distribution function:
$f_{\bf p}(t)=n_{\bf p}+f^{(1)}_{\bf p}e^{-i\omega t}+f^{(1)*}_{\bf p}e^{i\omega t}+f^{(2)}_{\bf p}$. 
Then,
\begin{gather}\label{KineticEq2}
  -i\omega f^{(1)}_{\bf p}+e{\bf E}\cdot\nabla_{\bf p}n_{\bf p}
  {-\omega_c\frac{\partial f^{(1)}_{\bf p}}{\partial\varphi_p}}
  =
  -\frac{f^{(1)}_{\bf p}}{\tau}+\sum_{\bf p'}W^{a}_{\bf p',p}f^{(1)}_{\bf p'},\\
  \nonumber
i\omega f^{(1)*}_{\bf p}+e{\bf E}^*\cdot\nabla_{\bf p}n_{\bf p}
{-\omega_c\frac{\partial f^{(1)*}_{\bf p}}{\partial\varphi_p}}
=
-\frac{f^{(1)*}_{\bf p}}{\tau}+\sum_{\bf p'}W^{a}_{\bf p',p}f^{(1)*}_{\bf p'},
\\
\nonumber
e{\bf E}\cdot \nabla_{\bf p}f^{(1)*}_{\bf p}+
e{\bf E}^*\cdot \nabla_{\bf p}f^{(1)}_{\bf p}
{-\omega_c\frac{\partial f^{(2)}_{\bf p}}{\partial\varphi_p}}
=
-\frac{f^{(2)}_{\bf p}}{\tau}+\sum_{\bf p'}W^{a}_{\bf p',p}f^{(2)}_{\bf p'},
\end{gather}
or
\begin{gather}
\label{EqFirst01}
\left\{
\frac{i\omega}{\omega_c}
+\frac{\partial }{\partial\varphi_{\bf p}}
-\frac{1}{\omega_c\tau}
\right\}f^{(1)}_{\bf p}
  =
  \frac{ev}{\omega_c}(E_x\cos\varphi_{\bf p}+E_y\sin\varphi_{\bf p})\frac{\partial n_{\bf p}}{\partial\varepsilon_p}
  -\frac{1}{\omega_c}\sum_{\bf p'}W^{a}_{\bf p',p}f^{(1)}_{\bf p'},
\\
\left\{
-\frac{i\omega}{\omega_c}
+\frac{\partial }{\partial\varphi_{\bf p}}
-\frac{1}{\omega_c\tau}
\right\}f^{(1)*}_{\bf p}
  =
  \frac{ev}{\omega_c}(E^*_x\cos\varphi_{\bf p}+E^*_y\sin\varphi_{\bf p})\frac{\partial n_{\bf p}}{\partial\varepsilon_p}
  -\frac{1}{\omega_c}\sum_{\bf p'}W^{a}_{\bf p',p}f^{(1)*}_{\bf p'},
\\
\label{MainEqf2}
\left\{
\frac{\partial }{\partial\varphi_{\bf p}}
-\frac{1}{\omega_c\tau}
\right\}
f^{(2)}_{\bf p}
=
\frac{e}{\omega_c}{\bf E}\cdot \nabla_{\bf p}f^{(1)*}_{\bf p}
+\frac{e}{\omega_c}{\bf E}^*\cdot \nabla_{\bf p}f^{(1)}_{\bf p}
-\frac{1}{\omega_c}\sum_{\bf p'}W^{a}_{\bf p',p}f^{(2)}_{\bf p'}.
\end{gather}

Let us, first, find the solution of  Eq.~\eqref{EqFirst01}. 
For that, we split the function $f$ into symmetric and asymmetric parts:
\begin{eqnarray}
f_{\bf p}^{(1)}=f_1^s({\bf p})+f_1^a({\bf p}).
\end{eqnarray}
This gives us two equations:
\begin{eqnarray}
\label{EqFirst021}
\left\{
\frac{i\omega}{\omega_c}
+\frac{\partial }{\partial\varphi_{\bf p}}
-\frac{1}{\omega_c\tau}
\right\}f_1^s({\bf p})
  =
  \frac{ev_{\bf p}}{\omega_c}(E_x\cos\varphi_{\bf p}+E_y\sin\varphi_{\bf p})\frac{\partial n_{\bf p}}{\partial\varepsilon_p},
  \\
  \label{EqFirst022}
  \left\{
\frac{i\omega}{\omega_c}
+\frac{\partial }{\partial\varphi_{\bf p}}
-\frac{1}{\omega_c\tau}
\right\}f_1^a({\bf p})
  =
  -\frac{1}{\omega_c}\sum_{\bf p'}W^{a}_{\bf p',p}f_1^s({\bf p'}).
\end{eqnarray}
Then, we find the solution of Eq.~\eqref{EqFirst021}, $f_1^s$, and after, substitute it in~\eqref{EqFirst022} to find $f_1^a$.

We can rewrite Eq.~\eqref{EqFirst021} as:
\begin{eqnarray}
    \frac{\partial f_1^{s}}{\partial \varphi_{\bf p}}+i\alpha f_1^{s}=Q(\varphi_{\bf p}),
\end{eqnarray}
where
\begin{eqnarray}
\alpha&=&\frac{\omega+i/\tau}{\omega_c},\\
Q(\varphi_{\bf p})&=&\frac{ev_{\bf p}}{\omega_c}(E_x\cos\varphi_{\bf p}+E_y\sin\varphi_{\bf p})\frac{\partial n_{\bf p}}{\partial\varepsilon_{p}}.
\end{eqnarray}
The general solution reads as:
\begin{eqnarray}
 f_1^{s}(p,\varphi)&=&-\int\limits_0^\infty d\psi e^{i\alpha\psi}Q(\varphi+\psi)\\
 &=&
 ev_{\bf p}\left\{
E_x\frac{\tau(\cos\varphi-\omega_c\tau\sin\varphi-i\omega\tau\cos\varphi)}{\omega_c^2\tau^2-(\omega\tau+i)^2}
+
E_y\frac{\tau(\omega_c\tau\cos\varphi+\sin\varphi-i\omega\tau\sin\varphi)}{\omega_c^2\tau^2-(\omega\tau+i)^2}
 \right\}
 \left(-\frac{\partial n_{\bf p}}{\partial\varepsilon_{p}}\right)
 \\
 &=&
{
 ev_{\bf p}(C_x(\varphi)E_x+C_y(\varphi)E_y)
 \left(-\frac{\partial n_{\bf p}}{\partial\varepsilon_{p}}\right)
 },
\end{eqnarray}
where 
\begin{align*}
    C_x(\varphi) &= \frac{\tau(\cos\varphi-\omega_c\tau\sin\varphi-i\omega\tau\cos\varphi)}{\omega_c^2\tau^2-(\omega\tau+i)^2}, \\
    C_y(\varphi) &= \frac{\tau(\omega_c\tau\cos\varphi+\sin\varphi-i\omega\tau\sin\varphi)}{\omega_c^2\tau^2-(\omega\tau+i)^2},
\end{align*}
thus, we introduced the coefficients $C_{x(y)}=C_{x(y)}(\varphi,\omega,\omega_c)$.
We will only explicitly mention the dependence on $\varphi$ in these coefficients, remembering that they also depend on $\omega$ and $\omega_c$.
Here, we can mention several important relations:
\begin{eqnarray}
\nonumber
&&\int\limits_0^\infty d\psi 
e^{i\alpha\psi}
\cos(\varphi+\psi)
=\omega_c C_x(\varphi),~~~~~\int\limits_0^\infty d\psi 
e^{-i\alpha^*\psi}
\cos(\varphi+\psi)
=\omega_c C_x^*(\varphi),\\
\nonumber
&&\int\limits_0^\infty d\psi 
e^{i\alpha\psi}
\sin(\varphi+\psi)
=\omega_c C_y(\varphi),~~~~~\int\limits_0^\infty d\psi 
e^{-i\alpha^*\psi}
\sin(\varphi+\psi)
=\omega_c C_y^*(\varphi);\\
\nonumber
&&\frac{\partial C_x(\varphi)}{\partial\varphi}=-C_y(\varphi),~~~~~
\frac{\partial C_y(\varphi)}{\partial\varphi}=C_x(\varphi),\\
\nonumber
&&\frac{\partial C_x^*(\varphi)}{\partial\varphi}=-C_y^*(\varphi),~~~~~
\frac{\partial C_y^*(\varphi)}{\partial\varphi}=C_x^*(\varphi).
\end{eqnarray}

Similarly (changing $\omega\rightarrow-\omega$ and $E\rightarrow E^*$) we find:
\begin{eqnarray}
 f_1^{s*}(p,\varphi)
 &=&
 ev_{\bf p}\left\{
E_x^*\frac{\tau(\cos\varphi-\omega_c\tau\sin\varphi+i\omega\tau\cos\varphi)}{\omega_c^2\tau^2-(\omega\tau-i)^2}
+
E_y^*\frac{\tau(\omega_c\tau\cos\varphi+\sin\varphi+i\omega\tau\sin\varphi)}{\omega_c^2\tau^2-(\omega\tau-i)^2}
 \right\}
 \left(-\frac{\partial n_{\bf p}}{\partial\varepsilon_{p}}\right)
 \\
 &=&
{
 ev_{\bf p}(C_x^*(\varphi)E_x^*+C_y^*(\varphi)E_y^*)
 \left(-\frac{\partial n_{\bf p}}{\partial\varepsilon_{p}}\right)
 },
\end{eqnarray}
where 
\begin{align*}
    C_x^*(\varphi) &= \frac{\tau(\cos\varphi-\omega_c\tau\sin\varphi+i\omega\tau\cos\varphi)}{\omega_c^2\tau^2-(\omega\tau-i)^2} ,\\
    C_y^*(\varphi) &= \frac{\tau(\omega_c\tau\cos\varphi+\sin\varphi+i\omega\tau\sin\varphi)}{\omega_c^2\tau^2-(\omega\tau-i)^2}.
\end{align*}

Next, let us deal with the asymmetric part (using the same approach):
\begin{eqnarray}
\nonumber
 f_1^{a}(p,\varphi)&=&-\int\limits_0^\infty d\psi e^{i\alpha\psi}
 (-1)\sum_{\bf p'}W^{a}_{\bf p',p(\varphi\rightarrow\varphi+\psi)}
 {
 ev_{\bf p'}(C_x(\varphi')E_x+C_y(\varphi')E_y)
 \left(-\frac{\partial n_{\bf p'}}{\partial\varepsilon_{p'}}\right)
 }\\
 &=&
 \sum_{\bf p'}
{
 ev_{\bf p'}(C_x(\varphi')E_x+C_y(\varphi')E_y)
 \left(-\frac{\partial n_{\bf p'}}{\partial\varepsilon_{p'}}\right)
 }
 \int\limits_0^\infty d\psi e^{i\alpha\psi}
 W^{a}_{\bf p',p(\varphi\rightarrow\varphi+\psi)},
\\
\nonumber
 f_1^{a*}(p,\varphi)&=&-\int\limits_0^\infty d\psi e^{-i\alpha^*\psi}
 (-1)\sum_{\bf p'}W^{a}_{\bf p',p(\varphi\rightarrow\varphi+\psi)}
{
 ev_{\bf p'}(C_x^*(\varphi')E_x^*+C_y^*(\varphi')E_y^*)
 \left(-\frac{\partial n_{\bf p'}}{\partial\varepsilon_{p'}}\right)
 }\\
&=&
 \sum_{\bf p'}
{
 ev_{\bf p'}(C_x^*(\varphi')E_x^*+C_y^*(\varphi')E_y^*)
 \left(-\frac{\partial n_{\bf p'}}{\partial\varepsilon_{p'}}\right)
 }
 \int\limits_0^\infty d\psi e^{-i\alpha^*\psi}
 W^{a}_{\bf p',p(\varphi\rightarrow\varphi+\psi)}. 
\end{eqnarray}

Let us keep these expressions as they are for now and focus on $f_\mathbf{p}^{(2)}$ given in Eq.~\eqref{MainEqf2}.
In~\eqref{MainEqf2}, we have $e{\bf E}\cdot \nabla_{\bf p}f^{(1)*}_{\bf p}$ and $e{\bf E}^*\cdot \nabla_{\bf p}f^{(1)}_{\bf p}$:
\begin{eqnarray}
\label{EqSmth07}
e{\bf E}\cdot \nabla_{\bf p}f^{(1)*}_{\bf p}
&=&
e{\bf E}\cdot 
\nabla_{\bf p}
\left\{
{
ev_{\bf p}(C_x^*(\varphi)E_x^*+C_y^*(\varphi)E_y^*)
 \left(-\frac{\partial n_{\bf p}}{\partial\varepsilon_{p}}\right)
 }
 \right\}
\\
\nonumber
&&+
\int\limits_0^\infty d\psi e^{-i\alpha^*\psi}
 \sum_{\bf p'}
{
 ev_{\bf p'}(C_x^*(\varphi')E_x^*+C_y^*(\varphi')E_y^*)
 \left(-\frac{\partial n_{\bf p'}}{\partial\varepsilon_{p'}}\right)
 }
 e{\bf E}\cdot 
\nabla_{\bf p}
W^{a}_{\bf p',p(\varphi\rightarrow\varphi+\psi)},
\\
e{\bf E}^*\cdot \nabla_{\bf p}f^{(1)}_{\bf p}
&=&
e{\bf E}^*\cdot \nabla_{\bf p}
\left\{
{
ev_{\bf p}(C_x(\varphi)E_x+C_y(\varphi)E_y)
 \left(-\frac{\partial n_{\bf p}}{\partial\varepsilon_{p}}\right)
 }
\right\}
\\
\nonumber
&&+
\int\limits_0^\infty d\psi e^{i\alpha\psi}
 \sum_{\bf p'}
{
 ev_{\bf p'}(C_x(\varphi')E_x+C_y(\varphi')E_y)
 \left(-\frac{\partial n_{\bf p'}}{\partial\varepsilon_{p'}}\right)
 }
 e{\bf E}^*\cdot 
\nabla_{\bf p}W^{a}_{\bf p',p(\varphi\rightarrow\varphi+\psi)}.
\end{eqnarray}

Splitting, as before, the function $f_\mathbf{p}^{(2)}$ into the symmetric and asymmetric parts, 
\begin{eqnarray}
f_{\bf p}^{(2)}=f_2^s({\bf p})+f_2^a({\bf p}),
\end{eqnarray}
and following~\eqref{MainEqf2}, we find:
\begin{eqnarray}
\label{Eqf2main01}
\left\{
\frac{\partial }{\partial\varphi_{\bf p}}
-\frac{1}{\omega_c\tau}
\right\}
{f^{s}_{2}}
&=&
\frac{e}{\omega_c}{\bf E}\cdot 
\nabla_{\bf p}
\left\{
{
ev_{\bf p}(C_x^*(\varphi)E_x^*+C_y^*(\varphi)E_y^*)
 \left(-\frac{\partial n_{\bf p}}{\partial\varepsilon_{p}}\right)
 }
 \right\}\\
 \nonumber
 &&+
 \frac{e}{\omega_c}{\bf E}^*\cdot 
 \nabla_{\bf p}
\left\{
{
ev_{\bf p}(C_x(\varphi)E_x+C_y(\varphi)E_y)
 \left(-\frac{\partial n_{\bf p}}{\partial\varepsilon_{p}}\right)
 }
\right\},
\end{eqnarray}
and
\begin{eqnarray}
\label{Eqf2main00}
\left\{
\frac{\partial }{\partial\varphi_{\bf p}}
-\frac{1}{\omega_c\tau}
\right\}
f^{a}_{2}
&=&
-\frac{1}{\omega_c}\sum_{\bf p'}W^{a}_{\bf p',p}
{f_{2}^{s}}\\
\nonumber
&&+
\frac{e}{\omega_c}{\bf E}\cdot 
\nabla_{\bf p}
\int\limits_0^\infty d\psi e^{-i\alpha^*\psi}
 \sum_{\bf p'}
{
 ev_{\bf p'}(C_x^*(\varphi')E_x^*+C_y^*(\varphi')E_y^*)
 \left(-\frac{\partial n_{\bf p'}}{\partial\varepsilon_{p'}}\right)
 }
W^{a}_{\bf p',p(\varphi\rightarrow\varphi+\psi)}\\
\nonumber
&&+
\frac{e}{\omega_c}{\bf E}^*\cdot 
\nabla_{\bf p}
\int\limits_0^\infty d\psi e^{i\alpha\psi}
 \sum_{\bf p'}
{
 ev_{\bf p'}(C_x(\varphi')E_x+C_y(\varphi')E_y)
 \left(-\frac{\partial n_{\bf p'}}{\partial\varepsilon_{p'}}\right)
 }
 W^{a}_{\bf p',p(\varphi\rightarrow\varphi+\psi)}.
\end{eqnarray}

This last equation will allow us to find the electric current density using $f_{2}^{a}$.
For that, we need to find $f_{2}^{s}$.

Furthermore, in Eq.~\eqref{Eqf2main00} we have:
\begin{eqnarray}
&&\nabla_{\bf p}
\left\{
{
ev_{\bf p}(C_x^*(\varphi)E_x^*+C_y^*(\varphi)E_y^*)
 \left(-\frac{\partial n_{\bf p}}{\partial\varepsilon_{p}}\right)
 }
 \right\}
 =
 \frac{e}{m}
 (C_x^*(\varphi)E_x^*+C_y^*(\varphi)E_y^*)
 \left(-\frac{\partial n_{\bf p}}{\partial\varepsilon_{p}}\right)
 \mathbf{e}_{\bf p}\\
 \nonumber
 &&
 ~~~~~~~+
 \frac{e}{m}
 \left(\frac{\partial C_x^*(\varphi)}{\partial \varphi}E_x^*+\frac{\partial C_y^*(\varphi)}{\partial \varphi}E_y^*\right)
 \left(-\frac{\partial n_{\bf p}}{\partial\varepsilon_{p}}\right)
 \mathbf{e}_{\varphi_{\bf p}}
 +
 ev_{\bf p}(C_x^*(\varphi)E_x^*+C_y^*(\varphi)E_y^*)
 \left(-\frac{\partial^2 n_{\bf p}}{\partial\varepsilon_{p}^2}\right)
 v_{\bf p}
 \mathbf{e}_{\bf p}\\
 \nonumber
 &&=
 (C_x^*(\varphi)E_x^*+C_y^*(\varphi)E_y^*)
 \left\{
 \frac{e}{m}
 \left(-\frac{\partial n_{\bf p}}{\partial\varepsilon_{p}}\right)
 +
 ev^2_{\bf p}
 \left(-\frac{\partial^2 n_{\bf p}}{\partial\varepsilon_{p}^2}\right)
 \right\}
 \mathbf{e}_{\bf p}
 +
 \frac{e}{m}
 \left(-C_y^*(\varphi)E_x^*
 +C_x^*(\varphi)E_y^*\right)
 \left(-\frac{\partial n_{\bf p}}{\partial\varepsilon_{p}}\right)
 \mathbf{e}_{\varphi_{\bf p}}.
\end{eqnarray}
Then,
\begin{eqnarray}
&&\frac{e}{\omega_c}{\bf E}\cdot 
\nabla_{\bf p}
\left\{
{
ev_{\bf p}(C_x^*(\varphi)E_x^*+C_y^*(\varphi)E_y^*)
 \left(-\frac{\partial n_{\bf p}}{\partial\varepsilon_{p}}\right)
 }
 \right\}\\
 \nonumber
&& ~~~~~
 =
  \frac{e}{\omega_c}(C_x^*(\varphi)E_x^*+C_y^*(\varphi)E_y^*)
 \left\{
 \frac{e}{m}
 \left(-\frac{\partial n_{\bf p}}{\partial\varepsilon_{p}}\right)
 +
 ev^2_{\bf p}
 \left(-\frac{\partial^2 n_{\bf p}}{\partial\varepsilon_{p}^2}\right)
 \right\}
\left[E_x\cos\varphi_\mathbf{p}+E_y\sin\varphi_\mathbf{p}\right]\\
 \nonumber
&&~~~~~~~+
\frac{e^2}{m\omega_c}
 \left(-C_y^*(\varphi)E_x^*
 +C_x^*(\varphi)E_y^*\right)
 \left(-\frac{\partial n_{\bf p}}{\partial\varepsilon_{p}}\right)
\left[-E_x\sin\varphi_\mathbf{p}+E_y\cos\varphi_\mathbf{p}\right].
\end{eqnarray}
Also,
\begin{eqnarray}
&&\nabla_{\bf p}
\left\{
{
ev_{\bf p}(C_x(\varphi)E_x+C_y(\varphi)E_y)
 \left(-\frac{\partial n_{\bf p}}{\partial\varepsilon_{p}}\right)
 }
\right\}\\
 \nonumber
 &&~~~~~=
 (C_x(\varphi)E_x+C_y(\varphi)E_y)
 \left\{
 \frac{e}{m}
 \left(-\frac{\partial n_{\bf p}}{\partial\varepsilon_{p}}\right)
 +
 ev^2_{\bf p}
 \left(-\frac{\partial^2 n_{\bf p}}{\partial\varepsilon_{p}^2}\right)
 \right\}
 \mathbf{e}_{\bf p}
 +
 \frac{e}{m}
 \left(-C_y(\varphi)E_x
 +C_x(\varphi)E_y\right)
 \left(-\frac{\partial n_{\bf p}}{\partial\varepsilon_{p}}\right)
 \mathbf{e}_{\varphi_{\bf p}}.
\end{eqnarray}
Then,
\begin{eqnarray}
&&\frac{e}{\omega_c}{\bf E}^*\cdot 
\nabla_{\bf p}
\left\{
{
ev_{\bf p}(C_x(\varphi)E_x+C_y(\varphi)E_y)
 \left(-\frac{\partial n_{\bf p}}{\partial\varepsilon_{p}}\right)
 }
 \right\}\\
 \nonumber
&& ~~~~~
 =
  \frac{e}{\omega_c}(C_x(\varphi)E_x+C_y(\varphi)E_y)
 \left\{
 \frac{e}{m}
 \left(-\frac{\partial n_{\bf p}}{\partial\varepsilon_{p}}\right)
 +
 ev^2_{\bf p}
 \left(-\frac{\partial^2 n_{\bf p}}{\partial\varepsilon_{p}^2}\right)
 \right\}
\left[E_x^*\cos\varphi_\mathbf{p}+E_y^*\sin\varphi_\mathbf{p}\right]\\
 \nonumber
&&~~~~~~~+
\frac{e^2}{m\omega_c}
 \left(-C_y(\varphi)E_x
 +C_x(\varphi)E_y\right)
 \left(-\frac{\partial n_{\bf p}}{\partial\varepsilon_{p}}\right)
\left[-E_x^*\sin\varphi_\mathbf{p}+E_y^*\cos\varphi_\mathbf{p}\right].
\end{eqnarray}

Again returning to Eq.~\eqref{Eqf2main00}, we have there:
\begin{eqnarray}
\nabla_{\bf p}
W^{a}_{\bf p',p}
&=&
\nabla_{\bf p}
\left(
-\frac{V_0}{\tau}\eta\frac{\mu v}{\Delta^2}[p_y(p_x'^2-p_y'^2)-p_y'(p_x^2-p_y^2)+2p_xp_x'p_y'-2p_x'p_xp_y]
\right)\\
&=&
-\eta\frac{\mu vV_0}{\tau\Delta^2}
[(p_x'^2-p_y'^2)\mathbf{e}_y
-p_y'(2p_x\mathbf{e}_x-2p_y\mathbf{e}_y)
+2p_x'p_y'\mathbf{e}_x
-2p_x'(p_y\mathbf{e}_x+p_x\mathbf{e}_y)]\\
&=&
-\eta\frac{\mu vV_0}{\tau\Delta^2}
[(-2p_xp_y'
+2p_x'p_y'
-2p_x'p_y)\mathbf{e}_x
+
(p_x'^2-p_y'^2
+2p_yp_y'-2p_xp_x')\mathbf{e}_y].
\end{eqnarray}
Thus,
\begin{eqnarray}
\frac{e}{\omega_c}{\bf E}\cdot 
\nabla_{\bf p}W^{a}_{\bf p',p}
=
\eta\frac{\mu vV_0}{\tau\Delta^2}
\frac{e}{\omega_c}[E_x(-2p_xp_y'
+2p_x'p_y'
-2p_x'p_y)
+
E_y(p_x'^2-p_y'^2
+2p_yp_y'-2p_xp_x')].
\end{eqnarray}
Furthermore,
\begin{eqnarray}
\frac{e}{\omega_c}{\bf E}\cdot 
\nabla_{\bf p}
W^{a}_{\bf p',p(\varphi\rightarrow\varphi+\psi)}
&=&
\eta\frac{\mu vV_0}{\tau\Delta^2}
\frac{e}{\omega_c}[E_x(-2p\cos(\varphi+\psi)p_y'
+2p_x'p_y'
-2p\sin(\varphi+\psi)p_x')\\
\nonumber
&&+
E_y(p_x'^2-p_y'^2
+2p\sin(\varphi+\psi)p_y'-2p\cos(\varphi+\psi)p_x')].
\end{eqnarray}
Therefore,
\begin{eqnarray}
\frac{e}{\omega_c}{\bf E}\cdot 
\nabla_{\bf p}\int\limits_0^\infty d\psi e^{-i\alpha^*\psi}
W^{a}_{\bf p',p(\varphi\rightarrow\varphi+\psi)}
&=&
\eta\frac{\mu vV_0}{\tau\Delta^2}\frac{e}{\omega_c}
\left[E_x\left(-2pp_y'C_x^*(\varphi)
+2p_x'p_y'\frac{i\omega_c\tau}{-\omega\tau+i}
-2pp_x'C_y^*(\varphi)\right)\right.\\
\nonumber
&&\left.+
E_y\left((p_x'^2-p_y'^2)\frac{i\omega_c\tau}{-\omega\tau+i}
+2pp_y'C_y^*(\varphi)-2pp_x'C_x^*(\varphi)\right)\right],
\end{eqnarray}
and
\begin{eqnarray}
\frac{e}{\omega_c}{\bf E^*}\cdot 
\nabla_{\bf p}\int\limits_0^\infty d\psi e^{i\alpha\psi}
W^{a}_{\bf p',p(\varphi\rightarrow\varphi+\psi)}
&=&
\eta\frac{\mu vV_0}{\tau\Delta^2}\frac{e}{\omega_c}
\left[E_x^*\left(-2pp_y'C_x(\varphi)
-2p_x'p_y'\frac{i\omega_c\tau}{-\omega\tau-i}
-2pp_x'C_y(\varphi)\right)\right.\\
\nonumber
&&\left.+
E_y^*\left(-(p_x'^2-p_y'^2)\frac{i\omega_c\tau}{-\omega\tau-i}
+2pp_y'C_y(\varphi)-2pp_x'C_x(\varphi)\right)\right].
\end{eqnarray}

Summing up yields
\begin{eqnarray}
\label{Eqf2main01}
\left\{
\frac{\partial }{\partial\varphi_{\bf p}}
-\frac{1}{\omega_c\tau}
\right\}
{f^{s}_{2}}
&=&
e(C_x^*(\varphi)E_x^*+C_y^*(\varphi)E_y^*)
 \left\{
 \frac{e}{m}
 \left(-\frac{\partial n_{\bf p}}{\partial\varepsilon_{p}}\right)
 +
 ev^2_{\bf p}
 \left(-\frac{\partial^2 n_{\bf p}}{\partial\varepsilon_{p}^2}\right)
 \right\}
\left[E_x\cos\varphi_\mathbf{p}+E_y\sin\varphi_\mathbf{p}\right]\\
 \nonumber
&&~~~~~~~+
\frac{e^2}{m}
 \left(-C_y^*(\varphi)E_x^*
 +C_x^*(\varphi)E_y^*\right)
 \left(-\frac{\partial n_{\bf p}}{\partial\varepsilon_{p}}\right)
\left[-E_x\sin\varphi_\mathbf{p}+E_y\cos\varphi_\mathbf{p}\right]
\\
 \nonumber
 &&+
 e(C_x(\varphi)E_x+C_y(\varphi)E_y)
 \left\{
 \frac{e}{m}
 \left(-\frac{\partial n_{\bf p}}{\partial\varepsilon_{p}}\right)
 +
 ev^2_{\bf p}
 \left(-\frac{\partial^2 n_{\bf p}}{\partial\varepsilon_{p}^2}\right)
 \right\}
\left[E_x^*\cos\varphi_\mathbf{p}+E_y^*\sin\varphi_\mathbf{p}\right]\\
 \nonumber
&&~~~~~~~+
\frac{e^2}{m}
 \left(-C_y(\varphi)E_x
 +C_x(\varphi)E_y\right)
 \left(-\frac{\partial n_{\bf p}}{\partial\varepsilon_{p}}\right)
\left[-E_x^*\sin\varphi_\mathbf{p}+E_y^*\cos\varphi_\mathbf{p}\right],
\end{eqnarray}
or in the short form,
\begin{eqnarray}
\label{Eqf2main01}
\left\{
\frac{\partial }{\partial\varphi_{\bf p}}
-\frac{1}{\omega_c\tau}
\right\}
{f^{s}_{2}}
&=&
Q_2(\varphi_\mathbf{p}),
\end{eqnarray}
and thus,
\begin{eqnarray}
\label{SecSynCor}
f_2^s=-\int\limits_0^\infty
d\psi e^{-\psi/\omega_c\tau}
Q_2(\varphi_\mathbf{p}+\psi).
\end{eqnarray}
%



\section{Skew scattering-mediated collision integral}

To account for the lack of inversion symmetry, we express the scattering probability in terms of the skew scattering mechanism, consistent with the kinetic theory of photogalvanic effects \cite{Belinicher1980}.
To derive the explicit form of the collision integral, we start from the expression, transforming the summation via integration:
\begin{equation}
\label{Eqint01}
    - \sum_{\mathbf{p}'} W_{\mathbf{p}',\mathbf{p}}^a f_2^s(\mathbf{p}') = - \frac{1}{(2\pi\hbar)^2} \int_0^\infty p' dp' \int_0^{2\pi} d\varphi' W_{\mathbf{p}',\mathbf{p}}^a f_2^s(\mathbf{p}'),
\end{equation}
%
%
%
where the anisotropic scattering probability is defined from Eq.~\eqref{EqAprob}:
\begin{equation}
\label{EqAprob2}
W_{\mathbf{p}',\mathbf{p}}^a = \frac{\eta \kappa V_0 v_p}{\Delta^2 \tau} \left[ p' p^2 \sin(2\varphi + \varphi') - p p'^2 \sin(\varphi + 2\varphi') \right] \delta(\varepsilon_p - \varepsilon_{p'}).
\end{equation}
To evaluate the integral Eq.~\eqref{Eqint01} explicitly, we can separate the angular-dependent part from the energy-dependent prefactors. 
%
Thus, first, we consider the integral over the scattering angle $\varphi'$. 
Substituting the scattering probability $W_{\mathbf{p}',\mathbf{p}}^a$ from Eq.~\eqref{EqAprob2}, we find
\begin{equation}
\begin{aligned}
\label{AngInt01}
& \int_0^{2\pi} d\varphi' \left[ p' p^2 \sin(2\varphi + \varphi') - p p'^2 \sin(\varphi + 2\varphi') \right] \left( \mathcal{C}_{\cos} \cos 2\varphi' + \mathcal{C}_{\sin} \sin 2\varphi' \right) \\
&= - p p'^2 \int_0^{2\pi} d\varphi' (\sin \varphi \cos 2\varphi' + \cos \varphi \sin 2\varphi') (\mathcal{C}_{\cos} \cos 2\varphi' + \mathcal{C}_{\sin} \sin 2\varphi') 
=
-\pi p p'^2 (\mathcal{C}_{\cos} \sin \varphi + \mathcal{C}_{\sin} \cos \varphi),
\end{aligned}
\end{equation}
where $\mathcal{C}_{\cos}$ and $\mathcal{C}_{\sin}$ are defined explicitly in terms of the electric field components $E_x$ and $E_y$ as:
\begin{align}
    \mathcal{C}_{\cos} &= (|E_x|^2 - |E_y|^2) K_1 + (E_y^* E_x + E_x^* E_y) \tau\omega_c K_2, \\
    \mathcal{C}_{\sin} &= (E_y^* E_x + E_x^* E_y) K_1 - (|E_x|^2 - |E_y|^2) \tau\omega_c K_2,
\end{align}
with
\begin{align*}
    K_1(\omega, \omega_c) &= 1 + \tau^2(\omega^2 - \omega_c^2) [1 + 2\tau^2\omega_c^2]
    ~~\textrm{and}~~ 
    K_2(\omega, \omega_c) = 3 + \tau^2(\omega^2 + 3\omega_c^2).
\end{align*}

Substituting Eq.~\eqref{AngInt01} back into the full integral Eq.~\eqref{Eqint01} and performing the transformation $p'dp' = m d\varepsilon'$, we utilize the property of the energy delta-function $\delta(\varepsilon_p - \varepsilon_{p'})$ which fixes $p' = p$ and $v_{p'} = v_p$. Grouping the velocity and momentum terms, we arrive at 
%
\begin{equation}
    -\sum_{\mathbf{p}'} W_{\mathbf{p}', \mathbf{p}}^a f_2^s(\mathbf{p}') = -\dfrac{\eta \kappa V_0 v_p^2 p^4 e^2 \tau \omega_c}{4\pi\hbar^2 \Delta^2 (1+4\tau^2\omega_c^2)\bigl[ (1+\tau^2(\omega_c^2-\omega^2))^2 + 4\tau^2\omega^2 \bigr]} \cdot\left( -\dfrac{\partial^2 n_{\mathbf{p}}}{\partial \varepsilon_p^2}\right)\bigl[ \mathcal{C}_{\cos} \sin \varphi + \mathcal{C}_{\sin} \cos \varphi \bigr].
\end{equation}

To find the asymmetric part of the second-order distribution function $f_2^a(\mathbf{p})$, we solve the linear first-order differential equation derived from the kinetic equation:
\begin{equation}
\left( \frac{\partial}{\partial \varphi_{\mathbf{p}}} - \frac{1}{\omega_c \tau} \right) f_2^a(\mathbf{p}) = - \frac{1}{\omega_c }\sum_{\mathbf{p}'} W_{\mathbf{p}',\mathbf{p}}^a f_2^s(\mathbf{p}')
=
-A \left[ \mathcal{C}_{\cos} \sin \varphi_{\mathbf{p}} + \mathcal{C}_{\sin} \cos \varphi_{\mathbf{p}} \right],
\end{equation}
where the prefactor $A$ reads as
\begin{equation}
A = \frac{\eta \kappa V_0 v_p^2 p^4 e^2 \tau }{4 \pi \hbar^2 \Delta^2 (1 + 4 \omega_c^2 \tau^2) \left[ (1 + \tau^2(\omega_c^2 - \omega^2))^2 + 4 \omega^2 \tau^2 \right]} \left( -\frac{\partial^2 n_{\mathbf{p}}}{\partial \varepsilon_p^2} \right).
\end{equation}
Accounting for the periodicity requirement for the distribution function with respect to the angle $\varphi_{\mathbf{p}}$ yields

\begin{equation}
\begin{split}
f_2^a(\mathbf{p}) &= -A e^{\frac{\varphi_{\mathbf{p}}}{\omega_c \tau}} \int \left[ \mathcal{C}_{\cos} \sin \varphi_{\mathbf{p}} + \mathcal{C}_{\sin} \cos \varphi_{\mathbf{p}} \right] e^{-\frac{\varphi_{\mathbf{p}}}{\omega_c \tau}} d\varphi_{\mathbf{p}} \\
&= \frac{A \omega_c \tau}{1 + \omega_c^2 \tau^2} \left[ (\mathcal{C}_{\cos} - \omega_c \tau \mathcal{C}_{\sin}) \sin \varphi_{\mathbf{p}} + (\omega_c \tau \mathcal{C}_{\cos} + \mathcal{C}_{\sin}) \cos \varphi_{\mathbf{p}} \right].
\end{split}
\end{equation}
Summing up, we arrive at the final explicit form of the distribution function:
\begin{equation}
\begin{split}
f_2^a(\mathbf{p}) = & \frac{\eta \kappa V_0  p^4 e^2 \tau^2 }{4 \pi \hbar^2 m^2\Delta^2 (1 + \omega_c^2 \tau^2) (1 + 4 \omega_c^2 \tau^2) \bigl[ (1 + \tau^2(\omega_c^2 - \omega^2))^2 + 4 \omega^2 \tau^2 \bigr]} 
                     \left( -\frac{\partial^2 n_{\mathbf{p}}}{\partial \varepsilon_p^2} \right) \\
& \times \Bigl\{ \bigl[ (|E_x|^2 - |E_y|^2) (K_1 + \omega_c^2 \tau^2 K_2) - (E_x E_y^* + E_y E_x^*) \omega_c \tau (K_1 - K_2) \bigr] \sin \varphi_{\mathbf{p}} \\
& \quad + \bigl[ (|E_x|^2 - |E_y|^2) \omega_c \tau (K_1 - K_2) + (E_x E_y^* + E_y E_x^*) (K_1 + \omega_c^2 \tau^2 K_2) \bigr] \cos \varphi_{\mathbf{p}} \Bigr\}.
\end{split}
\end{equation}

\section{Skew skaterring-mediated second-order current response}
The current density can be found from the standard formula:
\begin{equation}
\label{Current01}
\mathbf{j}^{\eta} = \frac{2e}{(2\pi\hbar)^2} \int \mathbf{v_p} f_{2}^{a}(\mathbf{p}) d^2\mathbf{p} = \frac{2e}{(2\pi\hbar)^2} \int_{0}^{\infty} \int_{0}^{2\pi} v_p \begin{pmatrix} \cos\varphi_\mathbf{p} \\ \sin\varphi_\mathbf{p} \end{pmatrix} f_{2}^{a}(\mathbf{p}) p \, dp \, d\varphi_\mathbf{p},
\end{equation}
where $v_x = (p/m)\cos\varphi_\mathbf{p}$ and $v_y = (p/m)\sin\varphi_\mathbf{p}$.
The angular part of the integral for the $x$-component of the current density is (writing $\varphi$ instead of $\varphi_\mathbf{p}$ for brevity)
\begin{equation}
\begin{split}
\int_0^{2\pi} \cos \varphi  \Bigl[ \bigl((|E_x|^2 - |E_y|^2) (K_1 + \omega_c^2 \tau^2 K_2) + (E_x E_y^* + E_y E_x^*) \omega_c \tau (K_1 - K_2)\bigr) \sin \varphi \\
+ \bigl((|E_x|^2 - |E_y|^2) \omega_c \tau (K_1 - K_2) - (E_x E_y^* + E_y E_x^*) (K_1 + \omega_c^2 \tau^2 K_2)\bigr) \cos \varphi \Bigr] d\varphi \\
= (|E_x|^2 - |E_y|^2) (K_1 + \omega_c^2 \tau^2 K_2) + (E_x E_y^* + E_y E_x^*) \omega_c \tau (K_1 - K_2) \int_0^{2\pi} \sin \varphi \cos \varphi \, d\varphi \\ 
+ \bigl((|E_x|^2 - |E_y|^2) \omega_c \tau (K_1 - K_2) - (E_x E_y^* + E_y E_x^*) (K_1 + \omega_c^2 \tau^2 K_2)\bigr) \int_0^{2\pi} \cos^2 \varphi \, d\varphi.
\end{split}
\end{equation}
Using the trigonometric identities $\sin\varphi \cos\varphi = \frac{1}{2}\sin(2\varphi)$ and $\cos^2\varphi = (1+\cos(2\varphi))/2$, we obtain:
\begin{equation}
\pi ((|E_x|^2 - |E_y|^2) \omega_c \tau (K_1 - K_2) - (E_x E_y^* + E_y E_x^*) (K_1 + \omega_c^2 \tau^2 K_2)).
\end{equation}
Similarly, for the $y$-component:
\begin{equation}
\pi ((|E_x|^2 - |E_y|^2) (K_1 + \omega_c^2 \tau^2 K_2) + (E_x E_y^* + E_y E_x^*) \omega_c \tau (K_1 - K_2)).
\end{equation}

The radial part of the integral involves the second derivative of the Fermi distribution function $n_p$. 
Given the parabolic dispersion $\varepsilon_p = p^2/2m$ and $v_p = p/m$, we have:
\begin{equation}
\int_0^{\infty} p  \left( \frac{p}{m} \right)  \left( \frac{p^6}{m^2} \right) \left( - \frac{\partial^2 n_p}{\partial \varepsilon_p^2} \right) dp = \int_0^{\infty} \frac{p^8}{m^3} \left( - \frac{\partial^2 n_p}{\partial \varepsilon_p^2} \right) dp
= - \frac{7 p_F^5}{m}.
\end{equation}

Substituting the explicit expression  for $f_2^a(\mathbf{p})$ in Eq.~\eqref{Current01} and using  $\mathbf{v}_{\mathbf{p}} = \mathbf{p}/m$, we perform the integration. 
Then, the analytical expressions for the components of the current density read as
\begin{equation}
\begin{split}
j_x^{\eta(Is)} = &-\frac{7 e^3 p_F^5 \eta \kappa V_0 \tau^2}{8 \pi^2 \hbar^4 m \Delta^2 (1 + \omega_c^2 \tau^2)(1 + 4\omega_c^2 \tau^2) \left[ (1 + \tau^2(\omega_c^2 - \omega^2))^2 + 4\omega^2 \tau^2 \right]} \\
&\times \left[ (|E_x|^2 - |E_y|^2)(K_1 + \omega_c^2 \tau^2 K_2) + (E_x E_y^* + E_y E_x^*) \omega_c \tau (K_1 - K_2) \right],
\end{split}
\end{equation}
\begin{equation}
\begin{split}
j_y^{\eta(Is)} = &-\frac{7 e^3 p_F^5 \eta \kappa V_0 \tau^2}{8 \pi^2 \hbar^4 m \Delta^2 (1 + \omega_c^2 \tau^2)(1 + 4\omega_c^2 \tau^2) \bigl[ (1 + \tau^2(\omega_c^2 - \omega^2))^2 + 4\omega^2 \tau^2 \bigr]} \\
&\times \bigl[ -(E_x E_y^* + E_y E_x^*)(K_1 + \omega_c^2 \tau^2 K_2) + (|E_x|^2 - |E_y|^2) \omega_c \tau (K_1 - K_2) \bigr]
\end{split}
\end{equation}
or
\begin{eqnarray}
\label{EqContrib1x}
j_x^{\eta(Is)} &=& \mathcal{J}_s\frac{1}{(1 + \omega_c^2 \tau^2)(1 + 4\omega_c^2 \tau^2)}
\frac{\left. (K_1 + \omega_c^2 \tau^2 K_2)P_{lin}+
\omega_c \tau (K_1 - K_2) P_{cross}
\right.}{[1 + \tau^2(\omega_c^2 - \omega^2)]^2 + 4\omega^2 \tau^2}  
,
\end{eqnarray}
\begin{eqnarray}
\label{EqContrib1y}
j_y^{\eta(Is)} &=& \mathcal{J}_s\frac{1}{(1 + \omega_c^2 \tau^2)(1 + 4\omega_c^2 \tau^2)}
\frac{\left. \left(\omega_c \tau (K_1 - K_2)\right) P_{lin}
-
(K_1 + \omega_c^2 \tau^2 K_2)P_{cross}
\right.}{[1 + \tau^2(\omega_c^2 - \omega^2)]^2 + 4\omega^2 \tau^2}  
,
\end{eqnarray}
where
\begin{align}
    \mathcal{J}_s &= -\frac{7 e^3 p_F^5 \eta  V_0 \kappa \tau^2}{8 \pi^2 \hbar^4 m \Delta^2} = -\eta \frac{7 e^3 p_F^5 V_0 w \tau^2}{8\sqrt{2} \pi^2 \hbar^4 \sqrt{m} \Delta^{3/2}}.
\end{align}

These expressions constitute the first contribution to the electric current density.


\section{Evaluation of the asymmetric resonant Hall contribution}

The asymmetric part of the second-order distribution function $f_2^a(\varphi)$ is governed by the kinetic equation in the presence of a magnetic field:

\begin{eqnarray}
\left\{
\frac{\partial }{\partial\varphi_{\bf p}}
-\frac{1}{\omega_c\tau}
\right\}
f^{a}_{2}
&=&
\frac{e}{\omega_c}{\bf E}\cdot 
\nabla_{\bf p}
\int\limits_0^\infty d\psi e^{-i\alpha^*\psi}
 \sum_{\bf p'}
{
 ev_{\bf p'}(C_x^*(\varphi')E_x^*+C_y^*(\varphi')E_y^*)
 \left(-\frac{\partial n_{\bf p'}}{\partial\varepsilon_{p'}}\right)
 }
W^{a}_{\bf p',p(\varphi\rightarrow\varphi+\psi)}\\
\nonumber
&&+
\frac{e}{\omega_c}{\bf E}^*\cdot 
\nabla_{\bf p}
\int\limits_0^\infty d\psi e^{i\alpha\psi}
 \sum_{\bf p'}
{
 ev_{\bf p'}(C_x(\varphi')E_x+C_y(\varphi')E_y)
 \left(-\frac{\partial n_{\bf p'}}{\partial\varepsilon_{p'}}\right)
 }
 W^{a}_{\bf p',p(\varphi\rightarrow\varphi+\psi)},
\end{eqnarray}
or shortly
\begin{equation}
\left( \frac{\partial}{\partial \varphi} - \frac{1}{\omega_c \tau} \right) f_2^a(\varphi) = \mathcal{I}_2^a(\varphi),
\label{eq:kinetic_f2a}
\end{equation}
where the source term $\mathcal{I}_2^a(\varphi)$ is determined by the field-induced transitions and asymmetric scattering:
\begin{equation}
\mathcal{I}_2^a(\varphi) = \frac{\mathcal{P}_0}{\omega_c \tau} \left[ \frac{\cos\varphi}{D^*} \left( \mathcal{Q}_1^* + \omega_c \tau \mathcal{Q}_2^* \right) + \frac{\cos\varphi}{D} \left( \mathcal{Q}_1 + \omega_c \tau \mathcal{Q}_2 \right) + \frac{\sin\varphi}{D^*} \left( \mathcal{Q}_2^* - \omega_c \tau \mathcal{Q}_1^* \right) + \frac{\sin\varphi}{D} \left( \mathcal{Q}_2 - \omega_c \tau \mathcal{Q}_1 \right) \right],
\end{equation}
with the prefactor $\mathcal{P}_0$ and the resonance denominators $D$ and $D^*$ defined as:
\begin{equation}
\mathcal{P}_0 = \frac{2 \pi \eta \kappa v_F^2 V_0 e^2 m^2 p_F^2 \tau}{(2\pi\hbar)^2 \Delta^2 [1 + (\omega_c\tau)^2]}, \quad 
\begin{cases} 
D = (\omega_c \tau)^2 - (\omega \tau + i)^2, \\ 
D^* = (\omega_c \tau)^2 - (\omega \tau - i)^2. 
\end{cases}
\end{equation}
The field-dependent factors $\mathcal{Q}_{1,2}$ capture the polarization state of the incident radiation:
\begin{equation}
\begin{aligned}
\mathcal{Q}_1 &= [E_x^*(1 - i\omega\tau) - E_y^*\omega_c\tau][E_x(1 - i\omega\tau) + E_y\omega_c\tau]+ [E_x^*\omega_c\tau + E_y^*(1 - i\omega\tau)][E_x\omega_c\tau - E_y(1 - i\omega\tau)], \\
\mathcal{Q}_2 &= [E_x^*(1 - i\omega\tau) - E_y^*\omega_c\tau][E_x\omega_c\tau - E_y(1 - i\omega\tau)]- [E_x^*\omega_c\tau + E_y^*(1 - i\omega\tau)][E_x(1 - i\omega\tau) + E_y\omega_c\tau],
\end{aligned}
\end{equation}
and
\begin{equation}
\begin{aligned}
\mathcal{Q}_{1}^* &= [E_x(1+i\omega\tau) - E_y\omega_c\tau][E_x^*(1+i\omega\tau) + E_y^*\omega_c\tau] + [E_x\omega_c\tau + E_y(1+i\omega\tau)][E_x^*\omega_c\tau - E_y^*(1+i\omega\tau)],\\
\mathcal{Q}_{2}^* &= [E_x(1+i\omega\tau) - E_y\omega_c\tau][E_x^*\omega_c\tau - E_y^*(1+i\omega\tau)] - [E_x\omega_c\tau + E_y(1+i\omega\tau)][E_x^*(1+i\omega\tau) + E_y^*\omega_c\tau].
\end{aligned}
\end{equation}
Using the polarization parameters $P_{lin} = |E_x|^2 - |E_y|^2$ and $P_{cross} = E_x E_y^* + E_y E_x^*$, we can rewrite these factors as
\begin{eqnarray}
Q_1 = P_{lin} \left[ (1 - i\omega\tau)^2 + (\omega_c\tau)^2 \right], \\
Q_2 = -P_{cross} \left[ (1 - i\omega\tau)^2 + (\omega_c\tau)^2 \right],
\end{eqnarray}
and
\begin{eqnarray}
Q_1^* = P_{lin} \left[ (1 + i\omega\tau)^2 + (\omega_c\tau)^2 \right], \\
Q_2^* = -P_{cross} \left[ (1 + i\omega\tau)^2 + (\omega_c\tau)^2 \right].
\end{eqnarray}

\begin{figure*}[t]
    \centering
\includegraphics[width=\textwidth,height=8cm, keepaspectratio]{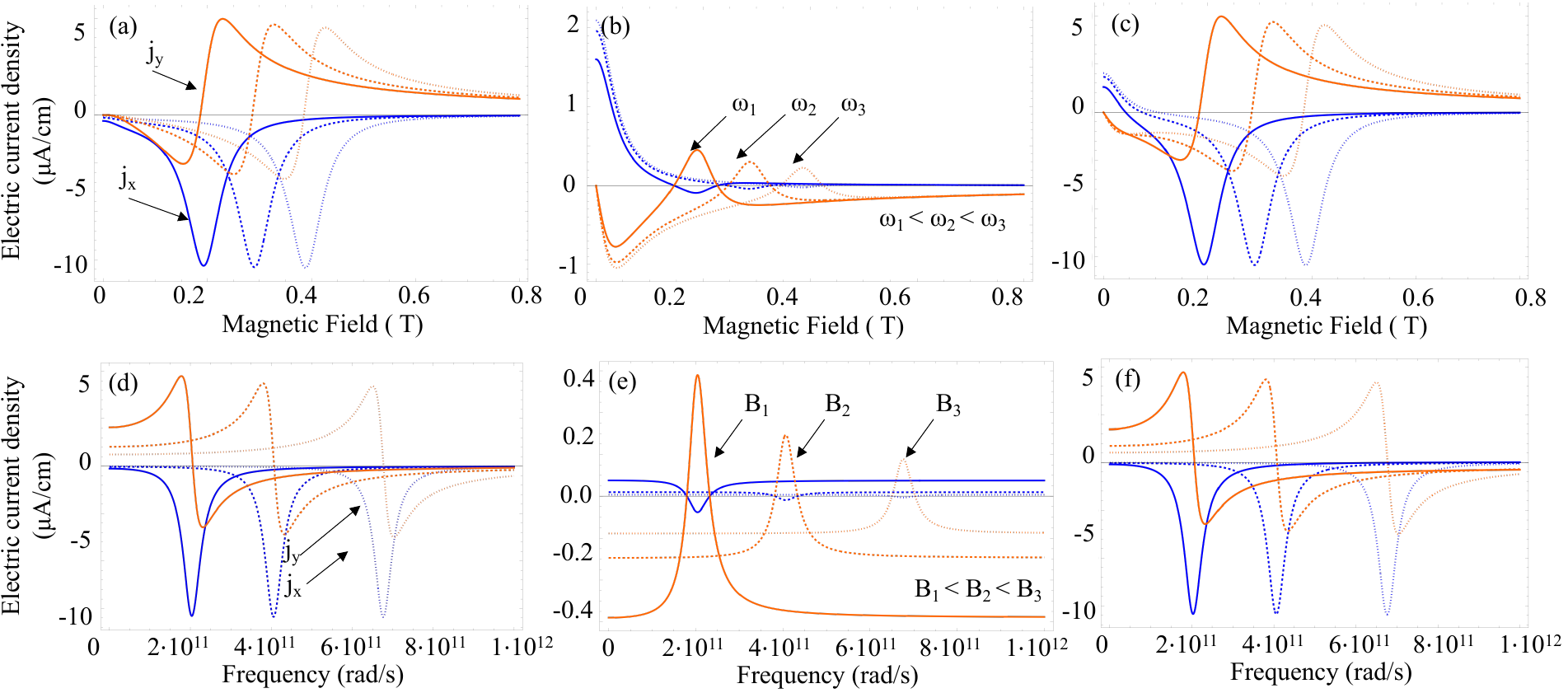}
    \caption{The cyclotron resonance via skew scattering mechanism: Electric current density as a function of the external magnetic field (a--c) and external light frequency (d--f). 
Blue and orange curves represent the longitudinal ($j_x$) and transverse ($j_y$) components, respectively. 
In (a--c), solid, dashed, and dotted lines correspond to $\omega_1\tau =5$, $\omega_2\tau=7.5$, and $\omega_3\tau=10$; 
in (d--f), the same styles denote magnetic field strengths 
$B_1 = 0.3~\text{T}$, 
$B_2 = 0.6~\text{T}$ and 
$B_3 = 1.0~\text{T}$.
Panels (a) and (d) correspond to the first contribution $j^{(I)}$, panels (b) and (e) correspond to the second contribution $j^{(II)}$, and panels (c) and (f) correspond to the total current density.}
\label{fig:resonanceSKEW}
\end{figure*}

The solution of the differential equation \eqref{eq:kinetic_f2a} is:
\begin{equation}
f_2^a(\varphi) = - \mathcal{P}_0 \left\{ \cos\varphi \left[ \frac{\tau}{D^*} \left( \mathcal{Q}_1^* + \omega_c \tau \mathcal{Q}_2^* \right) + \frac{\tau}{D} \left( \mathcal{Q}_1 + \omega_c \tau \mathcal{Q}_2 \right) \right] + \sin\varphi \left[ \frac{\tau}{D^*} \left( \mathcal{Q}_2^* - \omega_c \tau \mathcal{Q}_1^* \right) + \frac{\tau}{D} \left( \mathcal{Q}_2 - \omega_c \tau \mathcal{Q}_1 \right) \right] \right\}.
\end{equation}
%
Performing the angular integration, we arrive at the  expressions for the current densities:
\begin{align}
j_x^\text{$\eta(IIs)$} &= \frac{2\tilde{\cal J}_s}{1 + (\omega_c \tau)^2} \left[ \frac{\{1 + \tau^2(\omega_c^2 - \omega^2)\}\text{Re}(\mathcal{Q}_1 + \omega_c \tau \mathcal{Q}_2) + (2\omega \tau)\text{Im}(\mathcal{Q}_1 + \omega_c \tau \mathcal{Q}_2)}{[1 + \tau^2(\omega_c^2 - \omega^2)]^2 + 4\omega^2 \tau^2} \right], \\
j_y^\text{$\eta(IIs)$} &= \frac{2\tilde{\cal J}_s}{1 + (\omega_c \tau)^2} \left[ \frac{\{1 + \tau^2(\omega_c^2 - \omega^2)\}\text{Re}(\mathcal{Q}_2 - \omega_c \tau \mathcal{Q}_1) + (2\omega \tau)\text{Im}(\mathcal{Q}_2 - \omega_c \tau \mathcal{Q}_1)}{[1 + \tau^2(\omega_c^2 - \omega^2)]^2 + 4\omega^2 \tau^2} \right],
\end{align}
or
%
\begin{eqnarray}
&&j_x^{\eta(IIs)} = \frac{2\tilde{\cal J}_s \left(1 + \tau^2(\omega_c^2 - \omega^2)\right)^2 (P_{lin} - \omega_c \tau P_{cross})}{(1 + \omega_c^2 \tau^2)[1 + \tau^2(\omega_c^2 - \omega^2)]^2 + 4\omega^2 \tau^2}
-\frac{8\tilde{\cal J}_s \omega^2 \tau^2 (P_{lin} - \omega_c \tau P_{cross})}{(1 + \omega_c^2 \tau^2)([1 + \tau^2(\omega_c^2 - \omega^2)]^2 + 4\omega^2 \tau^2)},
\end{eqnarray}
\begin{eqnarray}
&&j_y^{\eta(IIs)} = -\frac{2\tilde{{\cal J}}_s \left(1 + \tau^2(\omega_c^2 - \omega^2)\right)^2 (P_{cross} + \omega_c \tau P_{lin})}{(1 + \omega_c^2 \tau^2)([1 + \tau^2(\omega_c^2 - \omega^2)]^2 + 4\omega^2 \tau^2)}
+\frac{8\tilde{\cal J}_s \omega^2 \tau^2 (P_{cross} + \omega_c \tau P_{lin})}{(1 + \omega_c^2 \tau^2)((1 + \tau^2(\omega_c^2 - \omega^2))^2 + 4\omega^2 \tau^2)},
\end{eqnarray}
where $\tilde{\cal J}_s=-{\cal J}_s/7$.

Plots illustrating the results regarding skew scattering are shown in Fig.~\ref{fig:resonanceSKEW}

%
%
%
%
%
%
%
%

\section{Warping-mediated contribution to electric current density}

When we consider the effect of the trigonal warping, the exact Boltzmann kinetic equation takes the form:
\begin{equation}
\frac{\partial f}{\partial t} + e \left( \mathbf{E}(t) + [\mathbf{V}_{\mathbf{p}} \times \mathbf{B}] \right) \cdot \nabla_{\mathbf{p}} f = -\frac{f - n(E_{\mathbf{p}})}{\tau},
\end{equation}
where $E_{\mathbf{p}} = \varepsilon_{\mathbf{p}} + \delta\varepsilon_{\mathbf{p}}$ is the total energy accounting for the warping $\delta\varepsilon_{\mathbf{p}}$, $n(E_{\mathbf{p}})$ is the equilibrium Fermi--Dirac distribution function, and $\mathbf{V}_{\mathbf{p}} = \nabla_{\mathbf{p}} E_{\mathbf{p}} = \mathbf{v}_\mathbf{p} + \delta\mathbf{v}_{\mathbf{p}}$ is the total electron group velocity ($\mathbf{v}_\mathbf{p}$ is the isotropic velocity, and $\delta\mathbf{v}_{\mathbf{p}}$ is the velocity correction due to warping).

We also expand the distribution function $f_w$ in powers of the electric field $\mathbf{E}$:

\begin{equation}
\frac{\partial f_{1w}}{\partial t} + e [\mathbf{v}_{\mathbf{p}} \times \mathbf{B}] \cdot \nabla_{\mathbf{p}} f_{1w} + \frac{f_{1w}}{\tau} = -e \mathbf{E}(t) \cdot \nabla_{\mathbf{p}} n(E_{\mathbf{p}}).
\end{equation}

\begin{equation}
\frac{\partial f_{2w}}{\partial t} + e [\mathbf{v}_{\mathbf{p}} \times \mathbf{B}] \cdot \nabla_{\mathbf{p}} f_{2w} + \frac{f_{2w}}{\tau} = -e \mathbf{E}(t) \cdot \nabla_{\mathbf{p}} f_{1w}.
\end{equation}

In the absence of trigonal warping ($\delta\varepsilon_{\mathbf{p}} = 0$), the solution to the first-order kinetic equation yields the linear distribution function:
\begin{equation}
f_{1w}^0(\mathbf{p}) = -n' \frac{e \tau}{m}\frac{(1 - i\omega\tau) (\mathbf{E} \cdot \mathbf{p}) + \omega_c\tau [\mathbf{E} \times \mathbf{p}]_z }{\left( 1 - i\omega\tau \right)^2 + (\omega_c\tau)^2},
\end{equation}
where $n' \equiv \partial n / \partial \varepsilon_{\mathbf{p}}$.
Then, we can evaluate the momentum-space gradient of $f_{1w}^0(\mathbf{p})$:
\begin{equation}
\begin{split}
\nabla_{\mathbf{p}} f_{1w}^0 = -\frac{e \tau}{m \left[ (1 - i\omega\tau)^2 + (\omega_c\tau)^2 \right]} & \left\{ n' \left[ (1 - i\omega\tau) \mathbf{E} + \omega_c\tau (\mathbf{e}_z \times \mathbf{E}) \right]  \right. \\
&+ \left. \frac{n'' \mathbf{p}}{m} \left[ (1 - i\omega\tau) (\mathbf{E} \cdot \mathbf{p}) + \omega_c\tau [\mathbf{E} \times \mathbf{p}]_z \right] \right\},
\end{split}
\end{equation}
where $n'' \equiv \partial^2 n / \partial \varepsilon_{\mathbf{p}}^2$.

To account for the trigonal warping of the energy spectrum, we expand the linear-in-field distribution function as $f_{1w} = f_{1w}^0 + \delta f_{1w}$. The governing equation for the first-order warping correction $\delta f_{1w}$ reads as
\begin{equation}
\frac{\partial \delta f_{1w}}{\partial t} + e [\mathbf{v}_\mathbf{p} \times \mathbf{B}] \cdot \nabla_{\mathbf{p}} \delta f_{1w} + \frac{\delta f_{1w}}{\tau} = -e [\delta\mathbf{v}_{\mathbf{p}} \times \mathbf{B}] \cdot \nabla_{\mathbf{p}} f_{1w}^0 - e \mathbf{E} \cdot \nabla_{\mathbf{p}} \left( n'(\varepsilon_{\mathbf{p}})\delta\varepsilon_{\mathbf{p}} \right),
\end{equation}
\begin{equation}
-2e \operatorname{Re} \left\{ \mathbf{E}^* \cdot \nabla_{\mathbf{p}} \delta f_1 \right\} = -2e \operatorname{Re} \left\{ E_x^* \frac{\partial \delta f_1}{\partial p_x} + E_y^* \frac{\partial \delta f_1}{\partial p_y} \right\}.
\end{equation}

Assuming a monochromatic field $\mathbf{E}(t) = \mathrm{Re}\{\mathbf{E}_0 e^{-i\omega t}\}$ yields:
\begin{equation}
\delta f_{1w}(\mathbf{p}, \varphi) = \frac{1}{\omega_c} \int_{0}^{\infty} d\theta \, e^{-\frac{1 - i\omega\tau}{\omega_c\tau} \theta} Q_{1w}(\varphi + \theta),
\end{equation}
where, 
\begin{equation}
\begin{split}
Q_{1w}(\varphi) = &-e \left\{ (\mathbf{E}_0 \cdot \mathbf{v}_\mathbf{p}) \, \delta\varepsilon_{\mathbf{p}} \, n'' + (\mathbf{E}_0 \cdot \delta\mathbf{v}_{\mathbf{p}}) n' \right\} \\
&- e \frac{\omega_c\tau}{(1 - i\omega\tau)^2 + \omega_c^2\tau^2} \left\{ [\mathbf{v}_\mathbf{p} \times \delta\mathbf{v}_{\mathbf{p}}]_z \left( (\mathbf{p} \cdot \mathbf{E}_0) - \omega_c\tau [\mathbf{p} \times \mathbf{E}_0]_z \right) n'' 
\right. \\
&\hspace{3.8cm} \left. 
+ 
\left( [\mathbf{E}_0 \times \delta\mathbf{v}_{\mathbf{p}}]_z - \omega_c\tau (\mathbf{E}_0 \cdot \delta\mathbf{v}_{\mathbf{p}}) \right) n' \right\}.
\end{split}
\end{equation}
Thus,
\begin{equation}
\begin{split}
\delta f_{1w} &= -\frac{e w p^4 n'' \tau}{2m} \Biggl\{ 
    E_x \operatorname{Re}\left[ \frac{e^{i4\varphi}}{1 - i(\omega + 4\omega_c)\tau} + \frac{e^{i2\varphi}}{1 - i(\omega + 2\omega_c)\tau} \right] \\
    &\hspace{2.3cm} + E_y \operatorname{Im}\left[ \frac{e^{i4\varphi}}{1 - i(\omega + 4\omega_c)\tau} - \frac{e^{i2\varphi}}{1 - i(\omega + 2\omega_c)\tau} \right] 
\Biggr\} \\
&\quad - 3e w p^2 n' \tau \Biggl\{ 
    E_x \operatorname{Re}\left[ \frac{e^{i2\varphi}}{1 - i(\omega + 2\omega_c)\tau} \right] - E_y \operatorname{Im}\left[ \frac{e^{i2\varphi}}{1 - i(\omega + 2\omega_c)\tau} \right] 
\Biggr\} \\
&\quad + \frac{3e w p^4 n'' \omega_c \tau^2}{2m \left[ (1 - i\omega\tau)^2 + \omega_c^2\tau^2 \right]} \Biggl\{ 
    (E_x + \omega_c\tau E_y) \operatorname{Im}\left[ \frac{e^{i4\varphi}}{1 - i(\omega + 4\omega_c)\tau} + \frac{e^{i2\varphi}}{1 - i(\omega + 2\omega_c)\tau} \right] \\
    &\hspace{4.5cm} + (E_y - \omega_c\tau E_x) \operatorname{Re}\left[ \frac{e^{i2\varphi}}{1 - i(\omega + 2\omega_c)\tau} - \frac{e^{i4\varphi}}{1 - i(\omega + 4\omega_c)\tau} \right] 
\Biggr\} \\
&\quad + \frac{3e w p^2 n' \omega_c \tau^2}{(1 - i\omega\tau)^2 + \omega_c^2\tau^2} \Biggl\{ 
    (E_x - \omega_c\tau E_y) \operatorname{Im}\left[ \frac{e^{i2\varphi}}{1 - i(\omega + 2\omega_c)\tau} \right] \\
    &\hspace{4.2cm} + (E_y + \omega_c\tau E_x) \operatorname{Re}\left[ \frac{e^{i2\varphi}}{1 - i(\omega + 2\omega_c)\tau} \right] 
\Biggr\}.
\end{split}
\end{equation}

Substituting the perturbation expansions $f_{1w} = f_{1w}^0 + \delta f_{1w}$ and $f_{2w} = f_{2w}^0 + \delta f_{2w}$ in the second-order kinetic equation yields
\begin{equation}
\frac{f_{2w}^0 + \delta f_{2w}}{\tau} + e \left[ (\mathbf{v}_\mathbf{p} + \delta\mathbf{v}_{\mathbf{p}}) \times \mathbf{B} \right] \cdot \nabla_{\mathbf{p}} (f_{2w}^0 + \delta f_{2w}) = -2e \operatorname{Re} \left\{ \mathbf{E}^* \cdot \nabla_{\mathbf{p}} (f_{1w}^0 + \delta f_{1w}) \right\}.
\end{equation}

Separating the contributions order-by-order in the trigonal warping parameter, we first isolate the unperturbed quadratic-in-field distribution function $f_2^0$:
\begin{equation}
\left( \frac{1}{\tau} + e [\mathbf{v}_\mathbf{p} \times \mathbf{B}] \cdot \nabla_{\mathbf{p}} \right) f_{2w}^0 = -2e \operatorname{Re} \left\{ \mathbf{E}^* \cdot \nabla_{\mathbf{p}} f_{1w}^0 \right\}.
\end{equation}

Transforming to momentum-space polar coordinates $(p, \varphi)$, this differential equation reduces to
\begin{equation}
\left( \frac{1}{\tau} - \omega_c \frac{\partial}{\partial \varphi} \right) f_{2w}^0(\varphi) = -2e \operatorname{Re} \left\{ \left( E^*_{x}\frac{\partial}{\partial p_x} + E^*_{y}\frac{\partial}{\partial p_y} \right) f_{1w}^0 \right\} \equiv Q_{2w}(\varphi).
\end{equation}
We find:
\begin{equation}
f_2^0(\varphi) = \frac{1}{\omega_c} \int_{0}^{\infty} d\theta \, e^{-\frac{\theta}{\omega_c \tau}} Q_2(\varphi + \theta),
\end{equation}
where the term $Q_2(\varphi+\theta)$ has the explicit form:
\begin{equation}
\begin{split}
Q_{2w}(\varphi+\theta) = \frac{2e^2 \tau}{m \left[ (1 - i\omega\tau)^2 + \omega_c^2 \tau^2 \right]} \operatorname{Re} \Biggl\{ 
& n' \left( E^*_{x} \left[ (1 - i\omega\tau) E_x - \omega_c\tau E_y \right] + E^*_{y} \left[ (1 - i\omega\tau) E_y + \omega_c\tau E_x \right] \right) \\
&+ \frac{n''}{m} (\mathbf{E}^* \cdot \mathbf{p}(\theta)) \Bigl[ (1 - i\omega\tau) (\mathbf{E} \cdot \mathbf{p}(\theta)) + \omega_c\tau [\mathbf{E} \times \mathbf{p}(\theta)]_z \Bigr] 
\Biggr\}.
\end{split}
\end{equation}

Evaluating the trajectory integral for $f_{2w}^0(\varphi)$ involves integrating the angular harmonics in $Q_{2w}(\varphi+\theta)$ over the cyclotron phase $\theta$:
\begin{equation}
f_{2w}^0(\varphi) = \frac{1}{\omega_c} \int_{0}^{\infty} d\theta \, e^{-\frac{\theta}{\omega_c \tau}} Q_{2w}(\varphi + \theta).
\end{equation}
We find:
\begin{equation}
\begin{split}
f_{2w}^0(\varphi) = & \frac{2e^2 \tau^2}{m \left[ (1 - i\omega\tau)^2 + \omega_c^2 \tau^2 \right]} \operatorname{Re} \Biggl\{ 
n' \Big( E^*_{x} \left[ (1 - i\omega\tau) E_x - \omega_c\tau E_y \right] \\
& + E^*_{y} \left[ (1 - i\omega\tau) E_y + \omega_c\tau E_x \right] \Big) + \frac{p^2 n''}{2m} \Biggl[ 
(1 - i\omega\tau) \Biggl( (E^*_{x}E_x + E^*_{y}E_y) \\
& + (E^*_{x}E_x - E^*_{y}E_y) \frac{\cos(2\varphi) - 2\omega_c\tau \sin(2\varphi)}{1 + 4\omega_c^2\tau^2} + (E^*_{x}E_y + E^*_{y}E_x) \frac{\sin(2\varphi) + 2\omega_c\tau \cos(2\varphi)}{1 + 4\omega_c^2\tau^2} \Biggr) \\
& + \omega_c\tau \Biggl( (E^*_{y}E_x - E^*_{x}E_y)  - (E^*_{x}E_y + E^*_{y}E_x) \frac{\cos(2\varphi) - 2\omega_c\tau \sin(2\varphi)}{1 + 4\omega_c^2\tau^2} \\
& + (E^*_{x}E_x - E^*_{y}E_y) \frac{\sin(2\varphi) + 2\omega_c\tau \cos(2\varphi)}{1 + 4\omega_c^2\tau^2} \Biggr)
\Biggr] \Biggr\}.
\end{split}
\end{equation}
The total warping-mediated nonlinear current density is evaluated by integrating the second-order distribution function over momentum space:
\begin{equation}
    \mathbf{j}^{\eta}_{warp} = e \sum_{\mathbf{p}} \mathbf{V}_{\mathbf{p}} f_{2w} = e \sum_{\mathbf{p}} \delta\mathbf{v}_{\mathbf{p}} f_{2w}^0 + e \sum_{\mathbf{p}} \mathbf{v}_{\mathbf{p}} \delta f_{2w}.
\end{equation}

\begin{figure*}[t]
    \centering
\includegraphics[width=\textwidth,height=8cm, keepaspectratio]{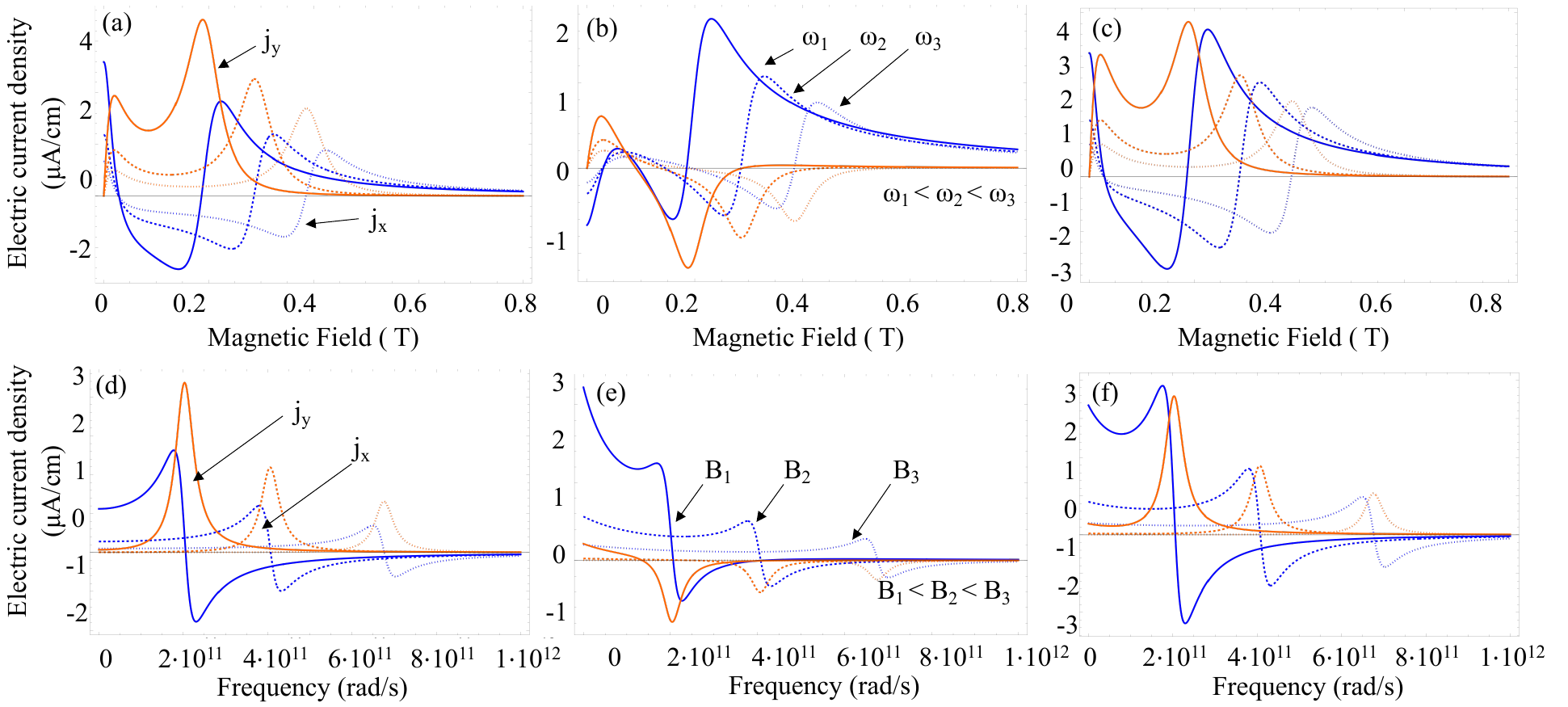}
    \caption{The cyclotron resonance via warping mechanism: Electric current density as a function of the external magnetic field (a--c) and external light frequency (d--f). 
Blue and orange curves represent the longitudinal ($j_x$) and transverse ($j_y$) components, respectively. 
In (a--c), solid, dashed, and dotted lines correspond to $\omega_1\tau =5$, $\omega_2\tau=7.5$, and $\omega_3\tau=10$; 
in (d--f), the same styles denote magnetic field strengths 
$B_1 = 0.3~\text{T}$, 
$B_2 = 0.6~\text{T}$ and 
$B_3 = 1.0~\text{T}$.
Panels (a) and (d) correspond to the first contribution $j^{(Iw)}$, panels (b) and (e) correspond to the second contribution $j^{(IIw)}$, and panels (c) and (f) correspond to the total current density.}
\label{fig:resonanceWARP}
\end{figure*}

The first contribution arises from the interplay between the velocity correction due to trigonal warping, $\delta\mathbf{v}_{\mathbf{p}}$, and the unperturbed second-order distribution function $f_{2w}^0$:
\begin{equation}
\mathbf{j}^{(Iw)} = e \sum_{\mathbf{p}} \delta\mathbf{v}_{\mathbf{p}} f_{2w}^0.
\end{equation}

%
%
%

The calculation gives:
\begin{equation}
\begin{split}
j_x^{\eta(Iw)} = & \frac{3 e^3 w \tau^2 p_F^2}{\pi\hbar^2 (1 + 4\omega_c^2\tau^2) \left[ (1 - \omega^2\tau^2 + \omega_c^2\tau^2)^2 + 4\omega^2\tau^2 \right]} \times \\
& \Biggl\{\left[ 1 + \omega^2\tau^2 + 3\omega_c^2\tau^2 + 2\omega_c^2\tau^2(\omega_c^2\tau^2 - \omega^2\tau^2) \right] (|E_x|^2-|E_y|^2 )+ \omega_c\tau \left[ 1 + 3\omega^2\tau^2 + \omega_c^2\tau^2 \right] (E^*_{x}E_y + E^*_{y}E_x)
\Biggr\}
\end{split}
\end{equation}
and
\begin{equation}
\begin{split}
j_y^{\eta(Iw)} = & \frac{3 e^3 w \tau^2 p_F^2}{\pi\hbar^2 (1 + 4\omega_c^2\tau^2) \left[ (1 - \omega^2\tau^2 + \omega_c^2\tau^2)^2 + 4\omega^2\tau^2 \right]}\times \\
& \Biggl\{\omega_c\tau \left[ 1 + 3\omega^2\tau^2 + \omega_c^2\tau^2 \right] (|E_x|^2-|E_y|^2)
- \left[ 1 + \omega^2\tau^2 + 3\omega_c^2\tau^2 + 2\omega_c^2\tau^2(\omega_c^2\tau^2 - \omega^2\tau^2) \right] (E^*_{x}E_y + E^*_{y}E_x)
\Biggr\},
\end{split}
\end{equation}
%
%
%
%
%
%
%

%
%
%
%
%
%

The second contribution to the warping-mediated electric current density arises from the unperturbed velocity combined with the second-order warping correction to the distribution function $\delta f_{2w}$:
\begin{equation}
\mathbf{j}^{\eta(IIw)} = e \sum_{\mathbf{p}} \mathbf{v}_
\mathbf{p} \delta f_{2w}.
\end{equation}
We find:
\begin{align}
j_x^{\eta(IIw)} ={}& - \frac{3 e^3 w \tau^2 p_F^2}{4 \pi \hbar^2} \Biggl[ (|E_x|^2-|E_y|^2)
\times \Biggl( \frac{1 - 2\omega_c^2\tau^2}{(1 + \omega_c^2\tau^2)\left[1 + (2\omega + 2\omega_c)^2\tau^2\right]} \nonumber \\
& \qquad - \frac{6\omega_c^2\tau^2 \left( 1 - \omega^2\tau^2 + \omega_c^2\tau^2 \right)}{\left[1 + (2\omega + 2\omega_c)^2\tau^2\right] \left[1 + (\omega - \omega_c)^2\tau^2\right] \left[1 + (\omega + \omega_c)^2\tau^2\right]} \Biggr) \nonumber \\
& - (E_x E_y^* + E_x^* E_y)
\times \Biggl( \frac{3\omega_c\tau}{(1 + \omega_c^2\tau^2)\left[1 + (2\omega + 2\omega_c)^2\tau^2\right]} \nonumber \\
& \qquad + \frac{3\omega_c\tau \left( 1 - 3\omega^2\tau^2 + \omega_c^2\tau^2 \right)}{\left[1 + (2\omega + 2\omega_c)^2\tau^2\right] \left[1 + (\omega - \omega_c)^2\tau^2\right] \left[1 + (\omega + \omega_c)^2\tau^2\right]} \Biggr) \Biggr],
\end{align}
\begin{align}
j_y^{\eta(IIw)} ={}& \frac{3 e^3 w \tau^2 p_F^2}{4 \pi \hbar^2} \Biggl[ (|E_x|^2-|E_y|^2)
\times \Biggl( \frac{3\omega_c\tau}{(1 + \omega_c^2\tau^2)\left[1 + (2\omega + 2\omega_c)^2\tau^2\right]} \nonumber \\
& \qquad + \frac{3\omega_c\tau \left( 1 - 3\omega^2\tau^2 + \omega_c^2\tau^2 \right)}{\left[1 + (2\omega + 2\omega_c)^2\tau^2\right] \left[1 + (\omega - \omega_c)^2\tau^2\right] \left[1 + (\omega + \omega_c)^2\tau^2\right]} \Biggr) \\
& + (E_x E_y^* + E_x^* E_y)
\times \Biggl( \frac{1 - 2\omega_c^2\tau^2}{(1 + \omega_c^2\tau^2)\left[1 + (2\omega + 2\omega_c)^2\tau^2\right]} \nonumber \\
& \qquad - \frac{6\omega_c^2\tau^2 \left( 1 - \omega^2\tau^2 + \omega_c^2\tau^2 \right)}{\left[1 + (2\omega + 2\omega_c)^2\tau^2\right] \left[1 + (\omega - \omega_c)^2\tau^2\right] \left[1 + (\omega + \omega_c)^2\tau^2\right]} \Biggr) \Biggr]. \nonumber
\end{align}
%
%
Plots illustrating the results regarding warping are shown in Fig.~\ref{fig:resonanceWARP}.

\section{Total net current calculation}

The photocurrent components in a single valley $\eta = \pm 1$ are described by:   

\begin{align}
j^{\eta}_{x} &= \chi_\eta(E_F^\eta) \left(|E_x|^2 - |E_y|^2\right) + \lambda_\eta(E_F^\eta) \left(E_x^* E_y + E_x E_y^*\right) \\
j^{\eta}_{y} &= -\chi_\eta(E_F^\eta) \left(E_x^* E_y + E_x E_y^*\right) + \lambda_\eta(E_F^\eta) \left(|E_x|^2 - |E_y|^2\right)
\end{align}

Due to time-reversal symmetry, the kinetic coefficients are odd in the valley index at zero magnetic field: $\chi_{\eta}(E_F) = \eta \chi_{\eta=1}(E_F)$ and $\lambda_\eta(E_F) = \eta \lambda_{\eta=1}(E_F)$, where $\chi_{\eta=1}, \lambda_{\eta=1}$ correspond to valley $\eta = +1$.   

An out-of-plane magnetic field $B_z$ couples to the orbital magnetic moment $\mathbf{m}_{\mathbf{p}} = (0, 0, \eta m_z)$, inducing a valley-dependent shift in the Fermi level:   

\begin{equation}
E_F^\eta = E_F + \eta m_z B_z
\end{equation}

Assuming $m_z B_z \ll E_F$, we expand $\chi_{\eta=1}(E_F^\eta)$ in a Taylor series:   

\begin{equation}
\chi_\eta(E_F^\eta) = \eta \, \chi_{\eta=1}(E_F + \eta m_z B_z) \approx \eta \, \chi_{\eta=1}(E_F) + m_z B_z \frac{\partial \chi_{\eta=1}(E_F)}{\partial E_F}
\end{equation}

Summing over both valleys ($\eta = \pm 1$), the zero-field valley-polarised terms cancel out, giving the total charge response coefficients:

\begin{align}
\chi &= \sum_{\eta = \pm 1} \chi_\eta(E_F^\eta) = 2 m_z B_z \frac{\partial \chi_{\eta=1}(E_F)}{\partial E_F} \\
\lambda &= \sum_{\eta = \pm 1} \lambda_\eta(E_F^\eta) = 2 m_z B_z \frac{\partial \lambda_{\eta=1}(E_F)}{\partial E_F}
\end{align}

The response consists of skew-scattering ($s$) and trigonal warping ($w$) contributions, $\chi_{\eta=1} = \chi_{\eta=1}^{(s)} + \chi_{\eta=1}^{(w)}$:   

\begin{itemize}
    \item Skew-scattering ($s$): $J_s \propto p_F^5 \propto E_F^{5/2} \implies \dfrac{\partial \chi_{\eta=1}^{(s)}}{\partial E_F} = \dfrac{5}{2 E_F} \chi_{\eta=1}^{(s)}$
    \item Trigonal warping ($w$): $J_w \propto p_F^2 \propto E_F^1 \implies \dfrac{\partial \chi_{\eta=1}^{(w)}}{\partial E_F} = \dfrac{1}{E_F} \chi_{\eta=1}^{(w)}$
\end{itemize}

\begin{align}
\chi &= \frac{m_z B_z}{E_F} \left[ 5 \chi_{\eta=1}^{(s)} + 2 \chi_{\eta=1}^{(w)} \right] \\
\lambda &= \frac{m_z B_z}{E_F} \left[ 5 \lambda_{\eta=1}^{(s)} + 2 \lambda_{\eta=1}^{(w)} \right]
\end{align}

where 
\begin{align}
\chi_{\eta=1}^{(s)} &= \chi^{(Is)} + \chi^{(IIs)}, \\
\chi_{\eta=1}^{(w)} &= \chi^{(Iw)} + \chi^{(IIw)}.
\end{align}

and for $\lambda_1$ is:

\begin{align}
\lambda_{\eta=1}^{(s)} &= \lambda^{(Is)} + \lambda^{(IIs)}, \\
\lambda_{\eta=1}^{(w)} &= \lambda^{(Iw)} + \lambda^{(IIw)}.
\end{align}

The total charge current components are given by:
\begin{align}
j_{x} &= \chi \left(|E_x|^2 - |E_y|^2\right) + \lambda \left(E_x^* E_y + E_x E_y^*\right) \\
j_{y} &= -\chi\left(E_x^* E_y + E_x E_y^*\right) + \lambda \left(|E_x|^2 - |E_y|^2\right)
\end{align}

These relations are discussed in the main text of the manuscript.


\vspace{5pt}


\bibliography{bib}
\bibliographystyle{apsrev4-2}